\documentclass[openacc]{AArsproca_new}%%%%where rsproca is the template name

\usepackage{lipsum}
\usepackage{multicol}
\usepackage[normalem]{ulem}
\usepackage{comment}
\usepackage{dirtytalk}
\usepackage{bbm}
\usepackage{caption}
\usepackage{subcaption}
\usepackage{arydshln}
\usepackage{mathtools}
\usepackage{lipsum} 
\usepackage{xcolor}
\usepackage{gensymb}
\usepackage{dirtytalk}
\usepackage{graphicx}
\usepackage{url}
\jname{rsos}
\Journal{R Soc Open Sci\ }

\begin{document}

%%%% Article title to be placed here
%\title{Impacts of distal airway constriction on patterns of inhaled particle deposition in a whole-lung computational model}
\title{Non-local impact of distal airway constrictions on patterns of inhaled particle deposition}
%Airway constriction in a whole-lung particle-deposition model

\author{%%%% Author details
James D. Shemilt$^{1}$, Alex Horsley$^{2}$, Jim M. Wild$^{3}$, Oliver E. Jensen$^{1}$, Alice B. Thompson$^{1}$, Carl A. Whitfield$^{1,2}$}

%%%%%%%%% Insert author address here
\address{$^{1}$Department of Mathematics, and $^{2}$Division of Immunology, Immunity to Infection and Respiratory Medicine, University of Manchester, Manchester, UK, $^{3}$POLARIS, Imaging Sciences, Department of Infection, Immunity and
Cardiovascular Disease, University of Sheffield, Sheffield, UK\\
\vspace{6pt}
JDS, 0000-0002-9158-0930; AH, 0000-0003-1828-0058; \\JMW, 0000-0002-7246-866; OEJ, 0000-0003-0172-6578; 
\\ABT, 0000-0001-9558-1554; CAW, 0000-0001-5913-735X
}

%%%% Subject entries to be placed here %%%%
\subject{biophysics, medical physics, biomechanics}

%%%% Keyword entries to be placed here %%%%

%%%% Insert corresponding author and its email address}
\corres{James D. Shemilt\\
\email{james.shemilt@manchester.ac.uk}}

%%%% Abstract text to be placed here %%%%%%%%%%%%
\begin{abstract}
Airway constriction and blockage in obstructive lung diseases {cause ventilation heterogeneity and create barriers to effective drug deposition}. Established computational particle-deposition models have not accounted for these impacts of disease. {We present a new particle-deposition model that calculates ventilation based on the resistance of each airway, such that ventilation responds to airway constriction. The model incorporates distal airway constrictions representative of cystic fibrosis, allowing us to investigate the resulting impact on patterns of deposition. Unlike previous models, our model predicts how constrictions affect deposition in airways throughout the lungs, not just in the constricted airways. Deposition is reduced in airways directly distal and proximal to constrictions. When constrictions are clustered together, central-airways deposition can increase significantly in regions away from constrictions, but distal-airways deposition in those regions remains largely unchanged.} %This results in more uneven deposition and a significant disparity between the impact on central-airways and distal-airways deposition in regions not directly affected by constrictions. 
We use our model to calculate lung clearance index (LCI), a clinical measure of ventilation heterogeneity, after applying constrictions of varying severities in one lobe. We find an increase in LCI coinciding with {significantly reduced} deposition in the affected lobe. {Our results show how the model provides a framework for development of computational tools that capture the impacts of airway disease, which could significantly affect predictions of regional dosing.}

%We use our model to calculate lung clearance index (LCI), a clinical measure of ventilation heterogeneity, in lungs with constrictions of various severities localised to one lobe. We find an increase in LCI coinciding with a significant drop in deposition throughout the affected lobe. 
\end{abstract}
%%%%%%%%%%%%%%%%%%%%%%%%%%%

%%%%%%%%%% Insert the texts which can accomdate on firstpage in the tag "fmtext" %%%%%

\begin{fmtext}

\section{Introduction}

Inhaled therapies, including mucolytics and antibiotics, are commonly used to treat cystic fibrosis (CF) \cite{henke2007mucolytics,smith2022inhaled}. Recent evidence has also shown that inhaled gene therapies can halt or slow decline of lung function \cite{alton2015repeated}. Airway constriction or blockage caused by thickened mucus layers or mucus plugging can create barriers to the effective application of inhaled therapies in CF. Typically, it is desired for inhaled drugs to be deposited evenly throughout the lungs' conducting airways, including in the distal conducting airways. These small airways can be susceptible to constriction or blockage even in early disease \cite{tiddens2010cystic}, so delivering a sufficient dose to them, particularly in more diseased and poorly ventilated regions of the lungs, can be challenging. Barriers to effective application of inhaled therapies are not unique to CF; airway blockage can also alter patterns of inhaled particle deposition in chronic obstructive pulmonary disease (COPD) or severe asthma  \cite{darquenne_2012_aerosol,labiris2003pulmonary}. {Experimental studies have shown that, whilst deposition is evidently reduced in blocked airways due to a lack of ventilation through them, total lung deposition can be increased in patients with obstructive lung disease \cite{anderson1989effect}. Innovation in the design of inhalers and nebulisers, and how they are administered, can make drug delivery to the lungs more efficient \cite{laube2000targeting,de2015effects}, but ventilation heterogeneity in patients}

\end{fmtext}

%%%%%%%%%%%%%%% End of first page %%%%%%%%%%%%%%%%%%%%%

\maketitle

%\begin{multicols}{0}

%\section{Introduction}
\clearpage 

\noindent 
{ with obstructive lung disease is still a significant barrier to uniform drug deposition throughout the lungs, particularly in the diseased airways \cite{venegas2005self,venegas2023measuring}. Experimental techniques, such as gamma scintigraphy, can provide information on regional particle deposition in patients, including measures such as central-to-peripheral deposition ratio \cite{de2010lung,virchow2018lung}. Resolution is limited in imaging, and whilst new technologies are providing more detailed data on regional lung deposition, resolving particle deposition on the scale of individual small airways is beyond the capabilities of current experimental measurements. Computational particle-deposition modelling has the potential to provide detailed spatial information on patterns of local particle deposition in the small airways but, to do so, the heterogeneous ventilation induced by airway constriction must be accounted for.}

Early whole-lung particle-deposition models, such as the \say{trumpet} models (e.g., \cite{taulbee1975theory}) and \say{single-path} models (e.g., \cite{yeh1980models}), modelled the airway tree as a one-dimensional structure, with physical properties that vary by depth in the lungs but without any other spatial heterogeneity. Attempts have been made to use similar models to investigate the effects of bronchoconstriction, where a uniform reduction in radius was applied to all airways at certain depths, leading to increased deposition in those airways \cite{martonen2003silico}. More realistic, spatially heterogeneous patterns of airway constriction cannot be simulated with these models. 

Multiple-Path Particle Dosimetry (MPPD) models \cite{Anjilvel_1995_multiple,asgharian2001particle,miller2016improvements,asgharian_2022_mixing} are well-established computational tools that calculate regional particle deposition in an asymmetric airway tree. They have been shown to predict total deposition well in healthy adult lungs \cite{asgharian_2022_mixing}. The MPPD model assumes that the flow rate through each airway is proportional to the volume of lung subtended by that airway. This means that constricting an airway would not generally reduce the predicted flow rate through that airway since the volume of airways and acini subtended by it would not necessarily change. A more sophisticated ventilation model has previously been incorporated into MPPD \cite{asgharian2006airflow,asgharian2006prediction}, which took into account lung compliance and resistance as well as the volume and capacity of the region distal to each airway when calculating the flow through it. The resistance of each airway was not directly related to its diameter, suggesting that constricting an airway may not have induced a significant change to the resistance or flow rate. The more complex ventilation model did allow for more heterogeneous ventilation of the lungs, but did not have a significant impact on lobar deposition rates in healthy lungs, leading the authors to recommend continued use of the simpler uniform ventilation model in subsequent versions of MPPD \cite{asgharian2006airflow,asgharian2006prediction}. It has been acknowledged that using a uniform ventilation model is not likely to accurately predict deposition in diseased lungs \cite{asgharian2006prediction,hofmann2011modelling}.

Whole-lung models that combine three-dimensional computational fluid dynamics (CFD) simulations in the central airways with simpler one-dimensional models for the distal airways and acini have recently been developed \cite{oakes_aerosol_2017,poorbahrami2021whole,kuprat2021efficient}. The distal airways are treated either as several smaller trumpet \cite{oakes_aerosol_2017} or multiple-path models \cite{kuprat2021efficient}. These approaches provide models of deposition throughout the whole lung, including detailed patterns of deposition in the central airways. Air flow into each of the lungs' lobes was inferred from experimental measurements. However, within the distal airways, the same simple ventilation models used in traditional trumpet or MPPD models were employed, so the effects of ventilation heterogeneity induced by distal airway constrictions could not easily be explored. Grill et al. \cite{grill2023silico} have recently developed a patient-specific particle-deposition model which uses a more sophisticated ventilation model. They validated their model's outputs against experimental particle-deposition data from healthy subjects, showing good agreement, and they simulated one example of deposition in diseased lungs by increasing the stiffness of a region of lung to represent localised fibrosis. Whilst more sophisticated particle-deposition models are being developed, the impacts of distal airway constriction representative of obstructive lung disease on deposition at the whole-organ scale have not yet been investigated.

Other studies have used CFD to model particle deposition (see, e.g., the review by Longest \& Holbrook \cite{longest2012silico}). This is typically limited to modelling a relatively small number of airways, {and whilst relative ventilation of the lungs' five lobes may be inferred from patient-specific imaging in some models \cite{de2010validation,hajian2016functional}, the small airways are not generally modelled, exhalation is generally not captured, and neither is any ventilation heterogeneity beyond differences in lobar ventilation.}
Some attempts have been made to incorporate bronchoconstriction in CFD models, but they have largely focused on simulating particle deposition in only two or three generations of airways \cite{longest2006transport,farkas2007simulation}. 
%Longest et al. \cite{longest2006transport} simulated particle deposition in a double bifurcation geometry, before and after reducing the diameter of all the airways uniformly. They found that more particles deposited in the constricted geometry, and that the deposition became more localised at specific hotspots, generally near the carinas of the bifurcations. Farkas et al. \cite{farkas2007simulation} also modelled airway constriction in a three-dimensional particle-deposition model, but again limited to simulating only a double bifurcation geometry. 
Walenga \& Longest \cite{walenga2016current} explored the effects of airway constriction in a geometry representative of the whole lungs, in which each lobe was modelled by a single path of airways, with the other airways branching off that single path being truncated. They applied a uniform constriction to all airways in their model; thorough exploration of heterogeneous patterns of distal airway constriction are beyond the current capabilities of CFD models. 

Whole-lung models that can capture ventilation heterogeneity induced by airway constriction have been developed to simulate gas transport without particle deposition. Foy et al. \cite{foy2017modelling} investigated the effects of airway constriction on multiple-breath washout (MBW), a clinical measurement of ventilation heterogeneity. They showed that MBW indices are sensitive primarily to airway constriction severity corresponding to a reduction in radius of between $80\%$ and $90\%$. It has been shown that variability in simulated MBW indices is also elevated when airways are constricted by similar amounts \cite{whitfield2018modelling}.

%[\cite{longest2006transport} for CFD in double airway bifurcation geometries with uniformly constricted airways. \cite{walenga2016current} similar approach but using a stochastic individual pathway (SIP) model for the small airways - doesn't allow for any heterogeneity in the ditsal airways. \cite{farkas2007simulation} (re-read!) does CFD in double bifurcation geometries in which contrictions are applied to the downstream  airways, or a tumor-like constriction is applied in different locations. Can have a significant effect on local particle deposition patterns but limited to a small number of airways. See refs [31-36] in Grill, including Williams et al papers, for extension of CFD beyond the central airways, but without realistic ventilation models. Check what Williams et al. say about their ventilation model.] 

In this study, we present a whole-lung particle-deposition model, in which ventilation of the lungs is derived based on the resistance of each conducting airway and is driven by the expansion and contraction of individual acini. Transport of an inhaled gas and deposition of particles from that gas throughout the lungs are then calculated. {Unlike models such as MPPD \cite{asgharian2001particle} that assume uniform ventilation, our model is capable of making physics-based predictions of how airway constrictions throughout the lungs lead to non-uniform, heterogeneous ventilation and the impacts this has on patterns of particle deposition. This advancement in modelling inhaled drug delivery in diseased lungs lays the groundwork for future development of patient-specific particle deposition models that can capture the effects of airway disease. In this study, we demonstrate how the predictions of particle deposition by the model can be fundamentally altered by airway constriction; the predictions provide understanding and quantification of the physical mechanisms by which airway constriction and blockage impact patterns of deposition throughout the lungs. By simulating particle deposition in lungs in which patterns of small airway constriction representative of airway diseases such as CF have been applied, we establish that our model can make predictions that have biologic plausiblity, which match with existing understanding of how airway blockage reduces deposition to those blocked airways, and which provide detailed insight into how deposition in each individual airway in the lungs may be affected by localised airway constrictions. 
}

An overview of the model setup is provided in \S\ref{sec:methods}, with full details in the supplementary material. The model can predict significant changes to patterns of ventilation and particle deposition throughout the lungs when there is bronchoconstriction. We investigate these effects by applying spatially heterogeneous patterns of distal airway constrictions, representative of airway disease. We apply several qualitatively different patterns of constriction, including constricting airways in one or two lobes only, constricting a number of localised clusters of distal airways, or constricting a number of airways distributed randomly throughout the lungs. 
%The rest of the paper will be organised as follows. In §\ref{sec:methods}, we present the details of the model setup, with the ventilation model described in §\ref{sec:vent_equations} and the particle transport and deposition model described in §\ref{sec:transport}. The asymmetric airway network geometries on which we solve the model equations are described in §\ref{sec:geometries} and solution methods are discussed in §\ref{sec:solnmethods}. With minor modifications to the model, we are able to simulate multiple-breath washout (MBW), a clinical test used to quantify ventilation heterogeneity \cite{horsley2009lung}. In §\ref{sec:mbw}, we outline how this is done. In §\ref{sec:results}, we present results from the particle-deposition model, \dots [Finish once results and discussion sections are written.] 
We simulate associated clinical measures of ventilation heterogeneity (lung clearance index, LCI) and regional dosing (gamma scintigraphy). 

\section{Methods}\label{sec:methods}

%\end{multicols}
\begin{figure*}[!tp]
    \centering
    \includegraphics[width=\textwidth]{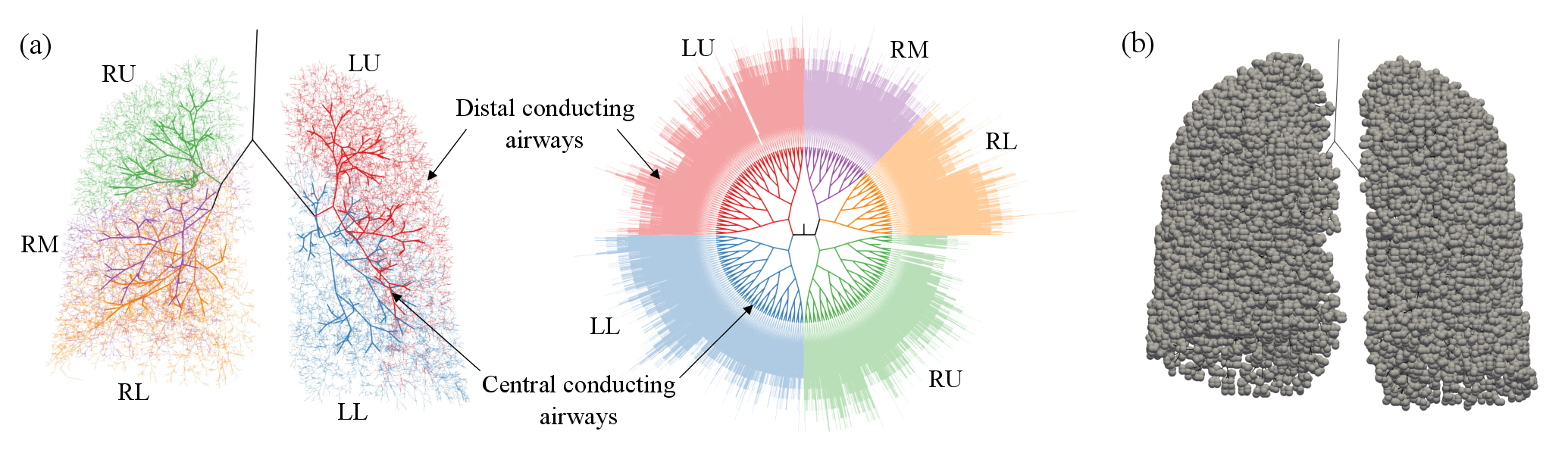}
    \caption{(a) Conducting airways in one lung network geometry, with the five lobes (Right Upper, Right Middle, Right Lower, Left Upper, Left Lower) highlighted. The network in 3D (left) is represented schematically as a planar graph (right). Path lengths from trachea (centre) to periphery are proportional to number of generations. (b) Acini, illustrated as spheres connected to the terminal conducting airways, for the same geometry.}
    \label{fig:geometry}
\end{figure*}
%\begin{multicols}{2}

\subsection{Lung networks}\label{sec:methods_network}

The lungs contain a large, asymmetric network of conducting airways, with acini attached to the terminal bronchioles. We model every airway individually, and every acinus. The conducting airways are treated as rigid tubes, while the acini expand and contract as the lungs are ventilated. %We derive equations describing ventilation through every airway, and transport of an inhaled gas and deposition of particles from that inhaled gas in every conducting and acinar airway in the network. %An overview of the model equations is provided below, with details in the supplementary material.
We use lung network geometries generated from computed tomography (CT) images of $n=4$ adolescents with CF but normal-range forced expiratory volume in one second. %The CT images were collected as part of a previous lung imaging study \cite{marshall2017detection}, which was approved by the National Research and Ethics Committee (REC 12/YH/0343). 
The CT images were collected clinically aligned with a previous lung imaging study \cite{marshall2017detection}; local research governance was in place to use anonymised CT scans retrospectively for imaging research purposes. 
The dimensions and positions of the largest central airways are taken directly from CT, and the remaining conducting airways are generated using a volume-filling algorithm \cite{tawhai2004ct}. Each of the four geometries has approximately $60,000$ conducting airways. 

Figure \ref{fig:geometry}(a) shows the conducting airways in one of the geometries. We denote the generation of an airway as one more than the number of bifurcations between it and the trachea, so that the trachea is denoted generation $1$. \say{Central conducting airways}, are those with generation less than $10$, and \say{distal conducting airways} are the remaining conducting airways (figure \ref{fig:geometry}a). The lungs are composed of five lobes which are indicated on figure \ref{fig:geometry}(a).  We also represent the geometry as a planar graph in figure \ref{fig:geometry}(a); we will present data from simulations below in this way to enable visualisation of deposition patterns across the entire network. We generate the planar graph representation of the network as follows. In each generation, the distal ends of the airways are placed on a circle with radius $(g-1)R_0$, where $g$ is the generation and $R_0$ is a constant spacing. Given the angular position, $\theta_p$, of the distal end of an airway in generation $g$, the angular positions of its daughter airways' distal ends are $\theta_p\pm2^{-g}\pi$, which ensures that all airways in the same generation have the same length and that daughter airways remain adjacent to each other and connected to their parent. 

To enable direct comparison of our model with MPPD, and to minimise differences purely due to lung size, we rescale the four lung geometries so that they are representative of an adult's lungs with functional residual capacity (FRC) of $3.3\mathrm{L}$. Given the measured FRC, $V_{\rm FRC}$, from each geometry, we uniformly increase the length and radius of each airway in that geometry by a factor of $(3.3\mathrm{L}/V_{\rm FRC})^{1/3}$. This results in a range of airway dead-space volumes between $101\mathrm{mL}$ and $145\mathrm{mL}$, which is realistic for adult lungs, noting we do not include the oral cavity \cite{hart1963relation}. {The original FRC values were between $1.1\mathrm{L}$ and $1.9\mathrm{L}$. We focus attention on simulating deposition in the scaled lung geometries so that we can compare directly to MPPD. We do not present results from simulations in the smaller, unscaled geometries, but we found that the main qualitative difference was higher inertial impaction in the central airways in the unscaled geometries, compared to in the scaled geometries, since the airway radii are smaller. Increased deposition in the central airways of children compared to adults has been highlighted in previous modelling studies (e.g., \cite{poorbahrami2021whole}). Investigating patterns of deposition in children or adolescents with CF is beyond the scope of this study.} In figure \ref{fig:Uncon}, we present data from simulations in all four {scaled} geometries and compare these to MPPD to validate the model. Subsequently, we present results from simulations in one of the four geometries, chosen arbitrarily, to focus on comparing deposition in the unconstricted geometry with deposition after applying constrictions to the distal airways. %We find that the impacts of constrictions in this geometry are at least qualitatively reproduced when constrictions are applied to other geometries. 

\subsection{Ventilation}\label{sec:methods_ventilation}

We assume that particle transport and deposition do not affect ventilation, so we first simulate ventilation and then use the calculated flow rates in each airway to simulate particle transport and deposition. The ventilation model is based on a previous model from Whitfield et al. \cite{whitfield2020spectral}. We provide an overview here and details in the supplementary material, §S2. We represent each lung geometry as a network of edges and vertices embedded in 3D space, with each edge representing an airway and vertices being placed at the ends of each edge. The resistance to air flow of each edge is calculated via Poiseuille's law, providing a linear relation between the difference in pressure along each airway and the flow rate through it. Whilst Poiseuille's law is not likely to accurately capture the complex dynamics in the largest central airways, it provides a computationally efficient simple approximation. {The underlying assumption in using Poiseuille's law is that flow in the airways is laminar, which is indeed the case in the smaller airways. However, in the largest central airways, there is likely to be turbulent flow, which is not captured by Poiseuille's law. We have also tested the nonlinear resistance model due to Pedley et al. \cite{pedley1970prediction}, which accounts for the formation of flow boundary layers at bifurcations, in simulations in healthy lung geometries but found it had minimal impact on total deposition. }

When simulating ventilation, each terminal vertex of the network represents an acinus. Figure \ref{fig:geometry}(b) illustrates these acini as individual spheres. We model each acinus as a viscoelastic bag, defining a relation between the pressure in the acinus, its volume and the pleural pressure. We assume that the breathing rate at the top of the trachea and the pleural pressure both vary sinusoidally in time, and that the pleural pressure is spatially uniform. We assume a breath time of $T_b=5\mathrm{s}$ and a tidal volume of $V_T=625\mathrm{mL}$. All acini are assumed to have the same initial volume and compliance. The total lung elastance, combining the contributions from all acini, is taken to be $6.82\mathrm{\,cmH_2OL}^{-1}$, which is derived from the measured value of $5\mathrm{\,cmH_2OL}^{-1}$ for lungs with FRC of $4.5\mathrm{L}$ \cite{harris2005pressure}, rescaled proportionally to the FRC of $3.3\mathrm{L}$. We tested the model after incorporating the effects of gravity on acinar dynamics by imposing a pleural pressure gradient and non-linear compliance (adapted from \cite{swan2012computational}); whilst this had some effect on the fraction of particles deposited in different lobes, the effects were relatively minor and the impact on total deposition was minimal. Incorporating these gravitational effects significantly increased computational time, so they were not used. 

\subsection{Particle transport and deposition}\label{sec:methods_transport}

We use the calculated air flow rates in each airway to solve an advection-diffusion equation for the transport of inhaled particles through the lungs. We solve for the concentration of inhaled particles throughout the lung network. Transport of particle concentration along an airway is governed by
\begin{equation}
    \frac{\partial \bar{c}}{\partial t} = \frac{\partial}{\partial z}\left(-\bar{u}\bar{c} + D_{\rm eff}\frac{\partial \bar{c}}{\partial z}\right) - s,
    \label{averageadvdiff}
\end{equation}
where $z$ is the axial coordinate, $\bar{c}(z,t)$ is the cross-sectionally averaged concentration, $\bar{u}(t)$ is the mean velocity, $D_{\rm eff}$ is an effective diffusivity that takes into account axial diffusion and dispersion, and $s$ represents loss due to particle deposition. To solve \eqref{averageadvdiff}, we first discretise each airway into several edges, thus defining a modified network. Then, we recast \eqref{averageadvdiff} using the machinery of discrete calculus, defining a discrete analogue of the advection-diffusion equation that can be efficiently solved on the network; this combines a finite difference approximation of \eqref{averageadvdiff} within each airway with mass conservation for inhaled gas and particles at bifurcations between airways. Full details are given in the supplementary material, \S{S3}.

We define $s$ in \eqref{averageadvdiff} by assuming that deposition occurs via three mechanisms: inertial impaction, gravitational sedimentation and diffusion. {We mostly present simulations of $4\mu\mathrm{m}$ diameter particles, as this is representative of the typical mass median aerodynamic diameter (MMAD) of particles generated by nebulisers used to administer inhaled therapeutics \cite{collins2009nebulizer}. For $4\mu\mathrm{m}$ diameter particles, the dominant deposition mechanisms are likely to be impaction and sedimentation \cite{hofmann2011modelling}. We also include deposition by diffusion in the model, which enables the model to describe deposition of smaller particles also. In figure \ref{fig:Uncon} below, we validate predictions against MPPD for a range of particle diameters from approximately $10^{-2}\mu\mathrm{m}$ to $8\mu\mathrm{m}$.} Following the approach of many other studies \cite{hofmann2011modelling}, we assume {that the three deposition} mechanisms act independently, and approximate deposition rates using semi-empirical or derived formulae. For impaction, we use a formula from Zhang et al. \cite{zhang_1997_impaction} derived from CFD simulations in airway bifurcation geometries. For sedimentation, we use a formula from Pich \cite{pich_1972_theory}, which was also used by Oakes et al. \cite{oakes_aerosol_2017}. For diffusion, we use a formula from Ingham \cite{ingham1991diffusion}, which has been validated against experimental results and CFD for sub-micron nanoparticles \cite{longest2007computational}. We do not model extrathoracic deposition, which is typically significant primarily for particles at least $6\mu\mathrm{m}$ in diameter \cite{darquenne_2016_total}, focusing instead on patterns of lung deposition. %\textcolor{black}{The particles in our simulations are assumed to be spherical, with density equal to that of water, so their geometric and aerodynamic diameters are equal. }
%\textcolor{black}{Whilst nebulisers produce particles with a range of diameters at once, we present simulations with monodisperse particles. Whilst this means our simulations do not precisely replicate a nebuliser therapy, our aim is to focus on understanding the physical mechanisms by which ventilation heterogeneity impacts particle deposition, so presenting data for single particle sizes allows us to isolate these effects. }

When simulating particle deposition, each acinus is modelled as a symmetric tree of airways. Each acinar airway consists of a duct of fixed radius, and a region of alveolar space with time-varying volume. The lengths and radii of the ducts are based on data from \cite{haefeli1988morphometry}, as is the relative volume of each airway within an acinar tree. The total volume of each acinus at each moment in time is taken from the ventilation simulation. 

We model particle deposition during a single breath, assuming that the initial concentration of particles is zero everywhere, and that the flux of concentration into the trachea is fixed during inhalation. To solve the discrete advection-diffusion equation, we approximate the time derivative using a first-order backwards Euler finite-difference scheme, and solve the resulting system of equations using the BiCGSTAB solver in the Eigen C++ library \cite{eigen_2010_v3}. {Using an implicit, backwards finite difference scheme provides improved stability of numerical solutions compared to many explicit schemes.} We use a time step of $\Delta t = 0.01\mathrm{s}$, which is small enough that total deposition is independent of the exact value. To discretise the airways to solve \eqref{averageadvdiff}, we ensure all edges within each airway have the same length, every airway, including in the acini, contains at least eight edges and edges are at most $200\mu\mathrm{m}$ long. The four discretised lung networks have between 2.5 and 3 million edges each. This discretisation ensures convergence of both conducting and acinar airway deposition; in the simulation presented in figure \ref{fig:Uncon}(a), approximately doubling the total number of edges changes the total deposition by only $0.2\%$. One typical simulation takes around two hours on a single-core $3.0\mathrm{GHz}$ processor.

In figure \ref{fig:Uncon}, we validate the model's output against MPPD simulations \cite{mppd_v3.04}. In the MPPD simulations, we use their \say{stochastic lung} geometry with an airway dead space of $113\mathrm{mL}$, which is within the range of dead-space values of our four lung geometries. Breath time, tidal volume and FRC are all the same as outlined in \S\ref{sec:methods_ventilation}, although MPPD uses a constant rate of inhalation and exhalation whilst we use a sinusoidal breathing profile. 

\subsection{Application of airway constrictions}\label{sec:methods_cons}

We investigate the effects of airway disease by applying several different patterns of 
constrictions to the distal conducting airways in one of the lung geometries. We always constrict a subset of the generation 12-15 airways only, and constrict these airways all with the same severity (a value between 0 and 1, the proportion by which we reduce the airways' radii). The way in which we select the airways for constriction, plus the constriction severity, determines the pattern of constriction. These choices are made both to illustrate key aspects of the physics of the system, and to represent typical features of CF disease. CF affects the small airways first \cite{tiddens2010cystic}, motivating our focus on distal airway constrictions. Ventilation MRI of patients with CF typically shows multiple small patches of poorly ventilated lung \cite{marshall2017detection,smith2018patterns}, motivating us to apply several localised clusters of airway constrictions in §\ref{sec:results_spatialcons}. We also investigate scenarios where constrictions are distributed randomly throughout the lungs or localised to one lobe. %Precise patterns of airway constriction applied are outlined in \S\ref{sec:results_LUvarcon} and \S\ref{sec:results_spatialcons}. 
We use the same breath time and tidal volume before and after applying constrictions. %; this results in a larger pleural pressure change during the breath after applying constrictions since the lungs' total resistance is increased. 
There is evidence that, in CF, the volume of air entering the lungs over a fixed time of several breaths does not change significantly as disease severity increases \cite{hart2002changes}, although breathing may become faster and shallower in very severe disease. Mild disease, which we focus on simulating here, is not likely to significantly affect breathing rates, and keeping breathing parameters fixed allows us to focus on the impacts of airway constriction.

\subsection{Simulated multiple-breath washout and gamma scintigraphy}\label{sec:methods_mbw}

We modify the model to also simulate multiple-breath washout (MBW). We simulate many breaths instead of one, set the deposition term in \eqref{averageadvdiff} to zero, set the diffusivity to the value for nitrogen gas in oxygen at $37\degree \mathrm{C}$, and assume an initially uniform concentration of tracer gas (nitrogen) throughout the lungs. 
We also assume that gas is well-mixed within each acinus, as the diffusivity of nitrogen is much higher than that of any sized aerosol we consider in the deposition model, so acinar mixing is likely to be much stronger. This MBW model is effectively equivalent to that of Foy et al. \cite{foy2017modelling}. %Further details are given in the supplementary material, \S{S5}. 
Lung clearance index (LCI), a clinical measure of ventilation heterogeneity, is then calculated as follows. Suppose that after $n_{\rm L}$ breaths, the concentration at the top of the trachea, $c_0(t)$, at end-exhalation is less than $1/40$ of the initial concentration, ${c}_0(n_{\rm L}T_b) < c_0(0)/40$, and that this is not the case for the previous breath. Then
\begin{equation}
    \mathrm{LCI} = \frac{V_{\rm ce}}{V_{\rm tr}}\left[{c}_0(0) - {c}_0(n_{\rm L}T_b)\right],\label{LCIdefn}
\end{equation}
where $V_{\rm ce}$ is the cumulative expired volume of gas and $V_{\rm tr}$ is the total volume of exhaled tracer gas \cite{horsley2009lung}. 

We use particle-deposition predictions to simulate gamma scintigraphy, a clinical measurement of regional dosing, which uses radiolabelling to image the distribution of particles in real patients. To do so, we divide the 2D plane into many pixels, sum the deposition that occurs within each pixel (across the whole depth of the lungs), and use kernel density estimation in \textsc{Matlab} to generate a deposition distribution that mimics a scintigraphy image. {The simulated scintigraphy is therefore derived entirely from the particle-deposition calculations, and is simply a method for visualising the simulation outcomes. Further details of the process of generating a simluated scintigraphy image are in the supplementary material, §S5.} %For scintigraphy, deposition that occurs at bifurcation points is assigned to the pixel in which the vertex of the network corresponding to that bifurcation point lies. When presenting data on single airway deposition (e.g., figure \ref{fig:Uncon}a below) or generational deposition (e.g., figure \ref{fig:Uncon}d below), we assign deposition that occurs at airway bifurcations to the parent airway or generation. 

\section{Results}\label{sec:results}
\subsection{Particle deposition in the lungs without airway constriction}\label{sec:results_Uncon}

%\end{multicols}
\begin{figure*}[!tp]
    \centering
    \includegraphics[width = \textwidth]{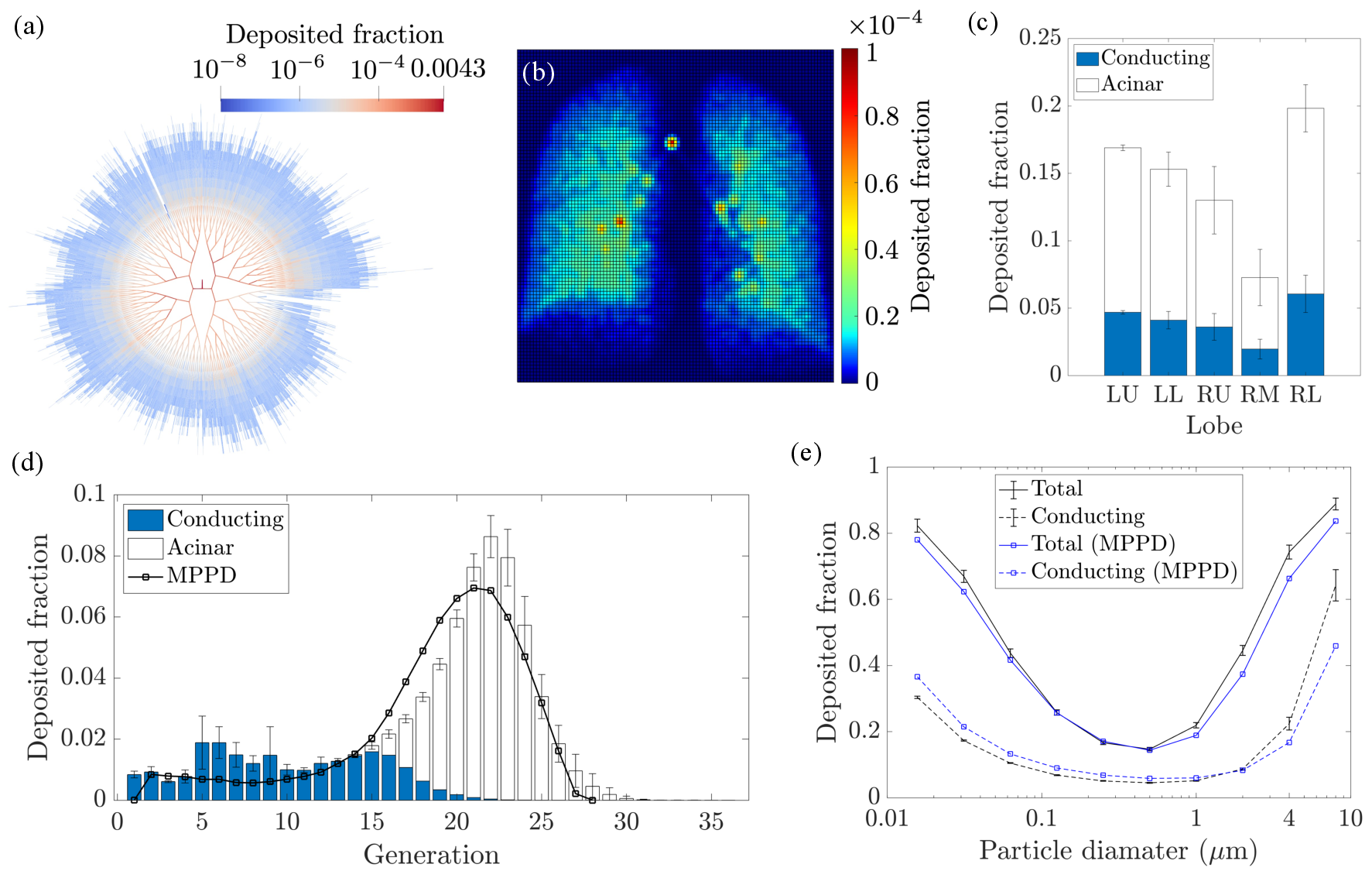}
    \caption{Data from simulations in the unconstricted lung geometries. In (a)-(d), particle diameter is $4\mu\mathrm{m}$. (a) Deposited fraction within each conducting airway in one geometry, shown as a planar graph. Deposition at bifurcations is assigned to the parent airway. (b) Simulated scintigraphy plot for the same example. (c) Deposited fraction in each lobe, showing mean $\pm$ one standard deviation across the four geometries. Deposited fraction separated into conducting-airways and acinar deposition. (d) Deposited fraction in each generation, showing mean $\pm$ one standard deviation across the four geometries, with conducting-airways and acinar deposition indicated. Generational deposition from MPPD is also shown. (e) Total and conducting-airways deposition, versus particle diameter, with comparison to MPPD. }
    \label{fig:Uncon}
\end{figure*}
%\begin{multicols}{2}

We first simulate particle deposition in all four lung geometries without applied bronchoconstriction, and compare results to MPPD for validation (figure \ref{fig:Uncon}). %Figure \ref{fig:Uncon}(a) shows the fraction of inhaled particles deposited in each conducting airway in one representative example simulation. 
Deposition is typically much higher in larger central airways than in smaller distal airways, but there is some heterogeneity in the spatial distribution of deposition due to asymmetries in the lung geometry (figure \ref{fig:Uncon}a). There is a range of path lengths between the trachea and the terminal conducting airways: in the geometry in figure \ref{fig:Uncon}(a), the shortest path terminates at  generation 8, and the longest paths extend beyond generation 20. Deposition is generally higher in regions where path lengths are longer (figure \ref{fig:Uncon}a). Since each terminal conducting airway connects to exactly one acinus, regions with a lot of long paths have a larger number of acini, so more particles are typically drawn through them and so more deposit. Simulated scintigraphy (figure \ref{fig:Uncon}b) demonstrates how the model can reproduce features of healthy deposition patterns: for example, deposition is relatively uniform spatially, with a few \say{hot spots} at central-airway bifurcation points where significant inertial impaction occurs. It shows higher deposition in the centre of the image where the lung is thicker, so the cumulative deposition across the whole depth of the lungs is higher. Figure \ref{fig:Uncon}(c) shows the distribution of deposited particles between the lungs' five lobes. 

To validate the model in healthy lung geometries, we compare deposition in the four lung geometries against results from MPPD (figure \ref{fig:Uncon}d,e). Deposition in each airway generation shows generally good agreement for $4\mu\mathrm{m}$ diameter particles (figure \ref{fig:Uncon}d), despite MPPD using a lung geometry based on different morphometric data. %Our simulations show higher deposition between generations 5 and 9 than MPPD, which is likely due to airways being wider on average in the MPPD model in these generations, leading to less inertial impaction, which is the dominant deposition mechanism in the central airways for $4\mu\mathrm{m}$ diameter particles. 
There is also good agreement with MPPD predictions of total deposition over a range of particle sizes, with only minor quantitative differences for particle diameters smaller than $0.1\mu\mathrm{m}$ or larger than $1\mu{\mathrm{m}}$. Higher conducting-airways deposition for large particles in our model is likely due to the central airways being narrower on average than in the MPPD lung geometry, causing higher impaction. We use different semi-empirical formulae than in MPPD to determine deposition rates, but despite this, agreement is still good.

{The MPPD model has been validated against experimental data in people with healthy lungs \cite{heyder1986deposition}, showing good agreement in total deposition over a range of particle sizes from $0.01\mu\mathrm{m}$ to $10\mu\mathrm{m}$ diameter particles \cite{asgharian2006prediction,asgharian_2022_mixing}. Good agreement (figure \ref{fig:Uncon}d,e) between our predictions of lung deposition and MPPD, therefore, also indicates that our model accurately captures lung deposition fractions across a wide range of particle sizes in healthy lungs. Comparison to the MPPD model provides validation in healthy lung geometries, but the MPPD model assumes a uniform ventilation so is not able to capture the effects of airway constriction or blockage on patterns of ventilation \cite{asgharian2006prediction,hofmann2011modelling}. In the following sections, we demonstrate our model's ability to predict heterogeneous patterns of ventilation and particle deposition after many airways in the lungs have been constricted. %Since we do not model extrathoracic deposition, instead focusing on investigating patterns of lung deposition, direct comparison with experimental data is more challenging, whilst validating our model against MPPD allows us to compare lung deposition fractions in isolation, without extrathoracic deposition. 
}

\subsection{Effects of varying airway constriction severity}\label{sec:results_LUvarcon}

%\end{multicols}
\begin{figure*}[!tp]
    \centering
    \includegraphics[width = \textwidth]{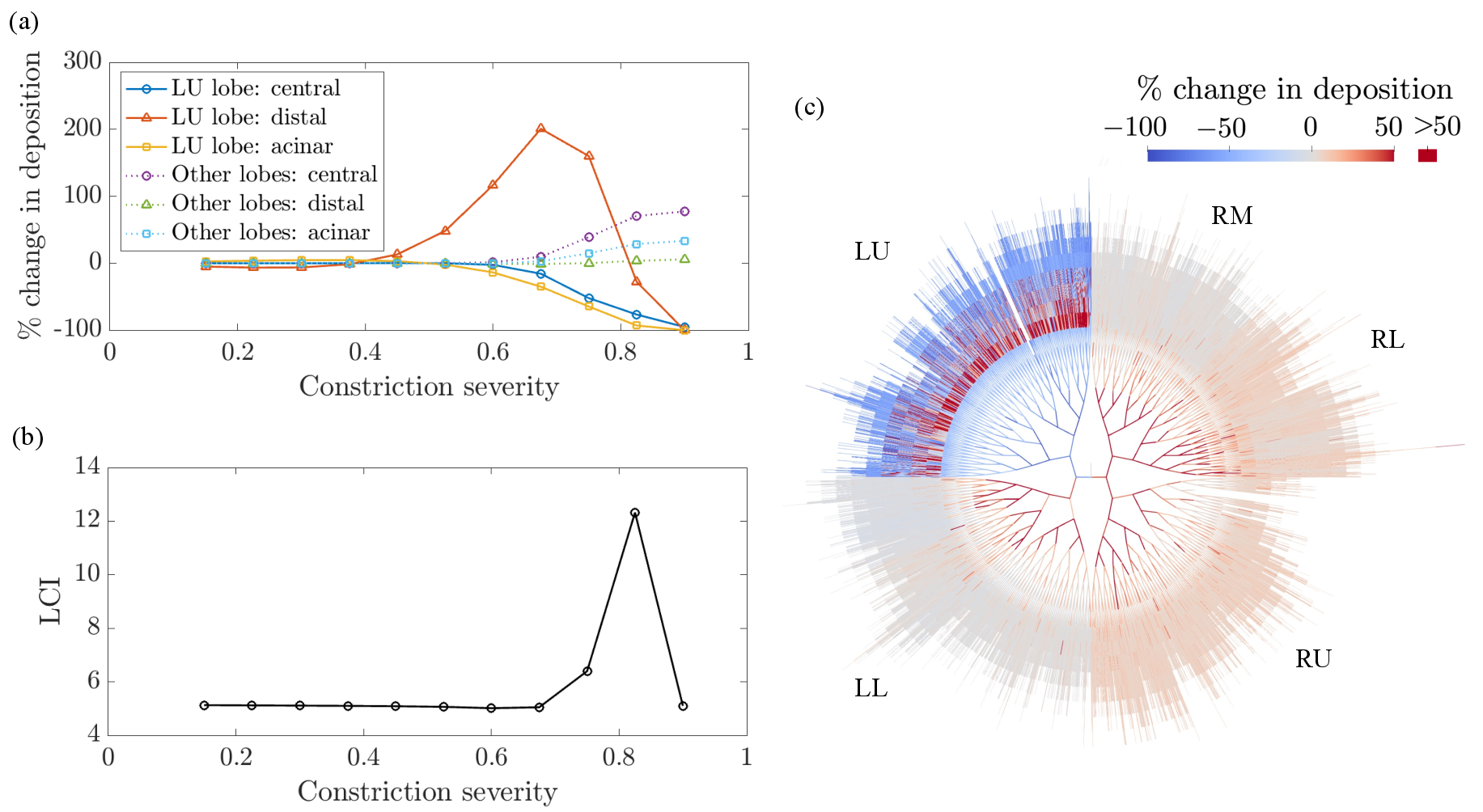}
    \caption{Simulations in which all generation 12-15 airways in the left upper lobe are constricted with varying severities (fraction by which airway radii are reduced). In each simulation, the constriction severity is the same for all left upper lobe airways. (a) Change in deposition in the affected lobe (LU), and the other lobes, separated into central conducting-airways deposition, distal conducting-airways deposition and acinar deposition. (b) Lung Clearance Index (LCI), from MBW simulations in the same constricted geometries. (c) Change in individual airway deposition in the simulation with constriction severity 0.825. }
    \label{fig:LUvarcons}
\end{figure*}
%\begin{multicols}{2}

To explore how the severity of airway constriction impacts deposition, we first constrict all generation 12-15 airways in a single lobe, with a range of constriction severities. The effect of increasing constriction severity on deposition in the affected airways is non-monotonic (figure \ref{fig:LUvarcons}a). {In these simulations, distal-airways deposition is not found to be} significantly impacted by mild constrictions but, when the severity exceeds 0.4, it increases as impaction is significantly enhanced in the narrowed airways. As the severity is increased further beyond 0.675, the increased resistance of the affected airways causes flow to be reduced into the affected lobe to such a degree that distal deposition decreases rapidly. Increased impaction in narrowed airways has been highlighted previously in simpler models \cite{martonen2003silico}, but these have not captured the subsequent decrease as airway resistance is further increased and ventilation patterns are altered. 

Central and acinar deposition in the affected lobe decrease as the constriction severity is increased and ventilation of the lobe is reduced (figure \ref{fig:LUvarcons}a). In the other lobes, deposition increases due to a higher proportion of the inhaled particles entering these lobes and the speed of air flow through them being faster. 
%Flow through the unconstricted lobes is increased since flow decreases through the constricted lobe and the flow rate into the trachea is assumed to be the same before and after applying constrictions. 
The strongest increase in deposition in the unconstricted lobes is in the central airways since impaction is the dominant deposition mechanism there, and this is enhanced by faster flow. By contrast, in the distal conducting airways, where sedimentation is the dominant mechanism, there is minimal change to deposition as shorter residence times for particles leads to reduced sedimentation, enough to largely offset any increased impaction. {The changes in deposition that we predict in airways that are not themselves constricted cannot be reproduced by existing models such as MPPD \cite{asgharian2001particle}, which assume uniform ventilation. 
Figure \ref{fig:LUvarcons}(a) shows that when the left upper lobe airways are severely constricted, we predict up to an $80\%$ increase in central-airways deposition in the rest of the lungs, and almost no change in distal-airways deposition. This quantification of the relative impact on deposition in the central and distal airways suggests a mechanism for how localised airway constriction can increase the central-peripheral deposition ratio throughout the lungs, a measure that can be increased \cite{de2010lung,virchow2018lung} in people with obstructive airway diseases, albeit with significant inter-subject variability in the experimental data.}

{As the constriction severity is increased in these simulations, other parameters such as particle size and breathing rate are kept constant, so as to isolate the impacts of airway constriction. The assumption of fixed tidal volume means that local changes in flow rate are solely due to redistribution of lobar ventilation. Some of the changes in regional deposition might be partially mitigated by adjusting these other parameters: for example, since impaction is weaker for smaller particles and at lower flow rates, the increase in central-airways deposition after applying severe constrictions may be smaller if the particle size or breathing rate was decreased. However, in the simulations shown in figure \ref{fig:LUvarcons}, we assume a typical flow rate for normal breathing and a particle size of $4\mu\mathrm{m}$, which is typical of the MMAD of many nebulised drugs. }

Figure \ref{fig:LUvarcons}(b) shows how lung clearance index (LCI) responds to the same applied constrictions. In agreement with previous modelling studies \cite{foy2017modelling,whitfield2018modelling}, we find that LCI is sensitive to a narrow range of constriction severities, which correspond to a reduction of airway radius of around $80\%$. Comparison with the deposition results (figure \ref{fig:LUvarcons}a) indicates that the increase in LCI coincides with where deposition is strongly reduced in the central and acinar airways in the constricted lobe, and increased elsewhere. When constriction severity is $0.825$, there is significantly reduced flow through the left upper lobe, but many of the constricted airways still receive increased deposition (figure \ref{fig:LUvarcons}c). The model predicts that when LCI is raised, particle deposition is likely to be more heterogeneous due to altered ventilation patterns, but constricted airways can still receive a significant dose. However, airways proximal or distal to those severely constricted airways are likely to receive a reduced dose, and a large fraction of the inhaled dose may be diverted to unconstricted regions of the lung. %Comparison of figures \ref{fig:LUvarcons}(a) and \ref{fig:LUvarcons}(b) indicates that LCI may not be likely to detect milder constrictions that can still drive increased deposition in the constricted airways, nor very severe constrictions that completely obstruct particle deposition. 
In keeping with previous modelling results, figure \ref{fig:LUvarcons}(b) suggests that LCI is mainly sensitive to severe (but not total) airway constrictions. Deposition of $4\mu\mathrm{m}$-diameter particles is, however, sensitive to a wider range of constriction severity, with contrasting effects from mild and severe airway constrictions.

\subsection{Effects of varying spatial patterns of airway constriction}\label{sec:results_spatialcons}

%\end{multicols}
\begin{figure*}[!tp]
    \centering
    \includegraphics[width=\textwidth]{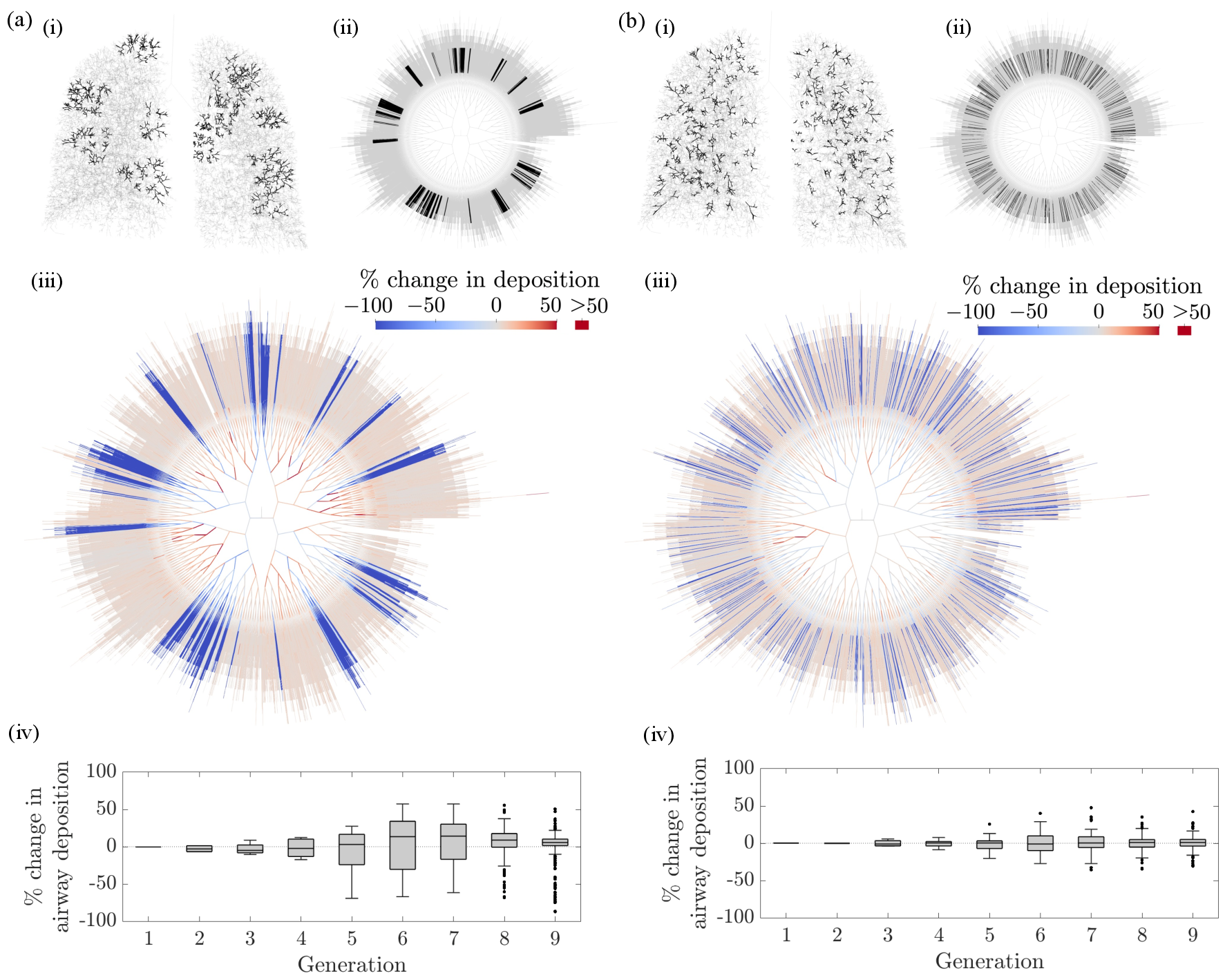}
\caption{Comparison of two patterns of distal airway constrictions. In (a), constrictions (severity 0.9) are applied in twelve clusters. Each cluster is generated by randomly selecting a generation-12 airway, then constricting all generation-12 airways within a radius of $R_C=2.4\mathrm{cm}$ of it and all of their descendants down to generation 15. We enforce that no clusters overlap. In total, 322 generation-12 airways, and all of their descendants down to generation 15, were constricted. In (b), 322 generation-12 airways were chosen at random, and they and their descendants down to generation 15 were constricted (severity 0.9). (i) Constricted airways highlighted in the 3D networks, and (ii) in the planar graph representations. (iii) Change in airway deposition versus deposition in the unconstricted geometry. (iv) Change in individual airway deposition as box plots for generations 1-9. 
    }
    \label{fig:PatchyComparison}
\end{figure*}
%\begin{multicols}{2}

\begin{figure*}[!tp]
    \centering
    \includegraphics[width = \textwidth]{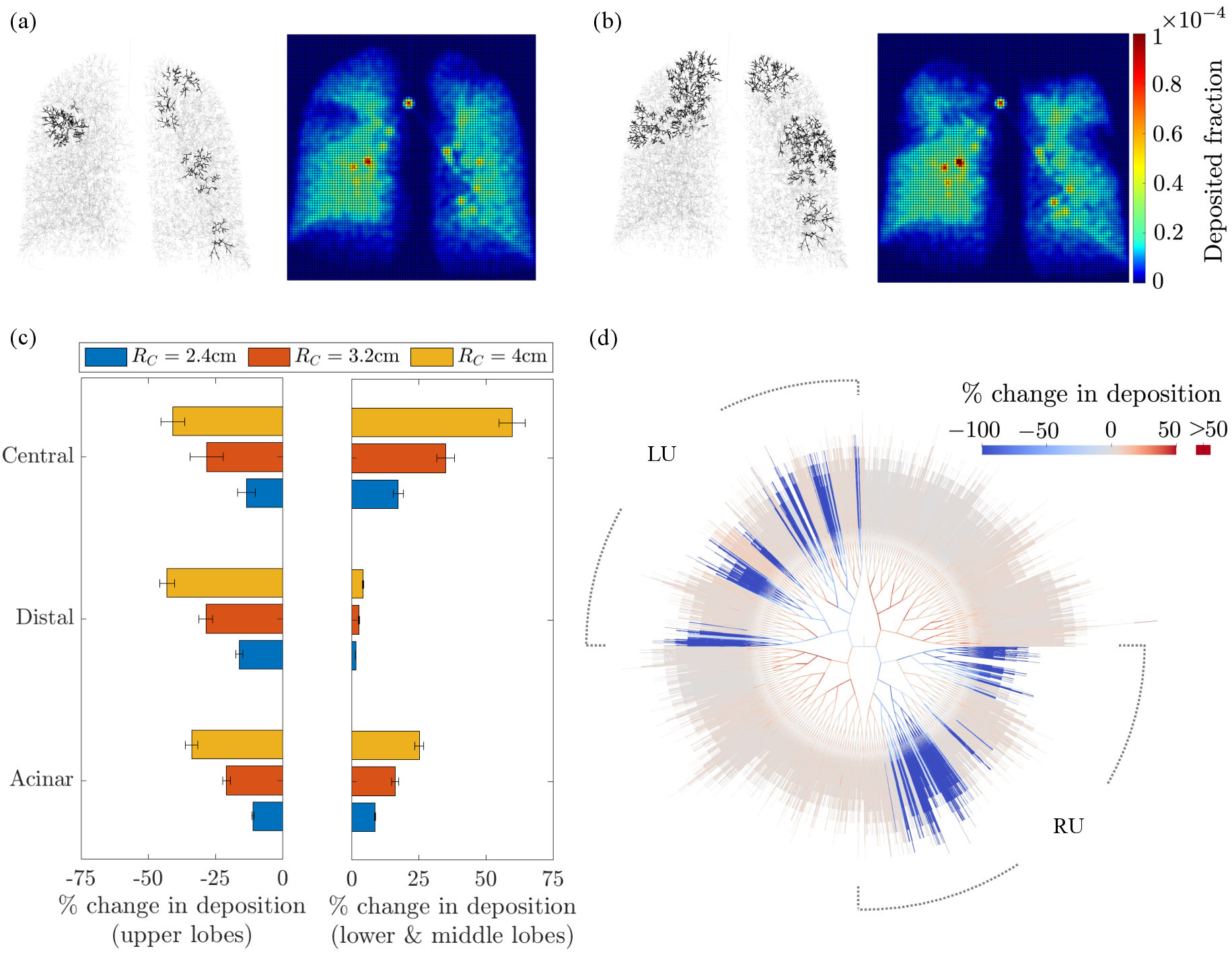}
    \caption{Data from simulations with six clusters of distal airway constriction (severity 0.9) in the upper lobes. In each simulation, six generation-12 airways were selected at random in the upper lobes, and all airways within a radius of $R_C$ of any of these were constricted, along with all of their descendants down to generation 15. Four realisations were simulated for each of $R_C=2.4\mathrm{cm}$, $R_C=3.2\mathrm{cm}$ and $R_C=4\mathrm{cm}$. (a) An example with $R_C=2.4\mathrm{cm}$, and (b) an example with $R_C=4\mathrm{cm}$, showing constricted airways (left) and {simulated} scintigraphy (right). (c) Percentage change in deposition versus the unconstricted case, showing change in the upper lobes (left), and in the other lobes (right). These data are also separated into central conducting, distal conducting and acinar deposition. Mean $\pm$ standard deviation across the four realisations is plotted for each case. (d) Change in individual airway deposition for an example with $R_C=3.2\mathrm{cm}$. 
    }
    \label{fig:PatchyCPA}
\end{figure*}

To investigate the impact of the spatial distribution of distal airway constrictions, we simulate two patterns of applied airway constrictions: clustered constrictions (figure \ref{fig:PatchyComparison}a) and constrictions distributed randomly throughout the lungs (figure \ref{fig:PatchyComparison}b). To generate clusters of constriction, we randomly select $N_C=12$ generation-12 airways, then constrict all generation-12 airways within a radius of $R_C=2.4\mathrm{cm}$ of any of these airways, and constrict all of their descendants down to generation 15. We apply severe constrictions (severity 0.9) to all of these airways. The same process is also used to generate clustered constrictions in figure \ref{fig:PatchyCPA}, but the number of clusters, $N_C$, and the radius, $R_C$, are different. We always enforce that no two clusters overlap. 

Both for clustered and randomly distributed constrictions, deposition is reduced significantly in the constricted airways, and in the airways distal to the constrictions, since flow through them is strongly reduced (figures \ref{fig:PatchyComparison}aiii,biii). Deposition is also reduced in the airways directly proximal to the constricted airways. This effect is stronger when multiple constricted airways are clustered together (figure
 \ref{fig:PatchyComparison}aiii), with decreases in deposition extending more proximally and the decreases being stronger than when constrictions are not clustered (figure
 \ref{fig:PatchyComparison}biii). This difference is also evidenced by comparison of figures \ref{fig:PatchyComparison}(aiv) and \ref{fig:PatchyComparison}(biv): more airways in generations 5-9 have significantly reduced deposition when constrictions are clustered than when they are not. 
 
 Both when the constrictions are clustered and when they are not, deposition typically increases in airways which are not directly proximal or distal to a constriction (figures \ref{fig:PatchyComparison}aiii,biii). This is because more of the inhaled gas is transported through these open paths once other paths are blocked, and flow through them is faster, so inertial impaction increases. When constrictions are clustered, increases in deposition in the central airways are generally larger than when constrictions are not clustered. This is evident in the comparison of figures \ref{fig:PatchyComparison}(aiii) and \ref{fig:PatchyComparison}(biii), and of figures \ref{fig:PatchyComparison}(aiv) and \ref{fig:PatchyComparison}(biv). Figure \ref{fig:PatchyComparison}(aiv) shows positive, although relatively small, median increases in airway deposition in several generations of central airways. Figure \ref{fig:PatchyComparison}(biv) shows almost exactly zero median change in the same generations when constrictions are not clustered. Variability in deposition in almost every generation of central airways is larger when constrictions are clustered. This highlights how clustered constrictions in the distal airways can cause more heterogeneous central-airways deposition.
 
%\end{multicols}
%\begin{multicols}{2}

To examine the impact of increasing the size of clusters of constriction, we now apply $N_C=6$ clusters to the upper lobes, and vary the cluster radius, $R_C$ (figure \ref{fig:PatchyCPA}). The process of generating clusters of constrictions is otherwise as described above. {Since $N_C$ is kept fixed, increasing $R_C$ typically leads to a marked increased in the number of airways being constricted, so that the simulations with higher values of $R_C$ are representative of more severely diseased states compared to those with lower values of $R_C$.} Simulated scintigraphy can be seen to clearly detect large clusters of constriction where many airways are blocked (figure \ref{fig:PatchyCPA}b). Where clusters of constrictions are smaller and fewer airways are constricted, not all constricted regions are visible on {simulated scintigraphy} (figure \ref{fig:PatchyCPA}a). In figure \ref{fig:PatchyCPA}(a), a patch of reduced deposition is clearly visible in the right lung, where two clusters are close together, but small clusters on the left lung, particularly those occurring in the centre of the image where lung depth is thickest, are not clear on the {simulated} scintigraphy. This highlights potential limitations of scintigraphy: it is more sensitive to severe blockages, and it may be more sensitive to constrictions that occur on the edges of the image field than those that occur in the centre. 

Figure \ref{fig:PatchyCPA}(c) quantifies how deposition changes in different regions of the lungs as the size of cluster of constrictions is increased. In the upper lobes, deposition decreases in the central, distal and acinar airways as the clusters are made larger and more airways are constricted. Elsewhere, deposition is increased as flow becomes faster through the unconstricted lobes. Central-airways deposition increases significantly, whilst there is almost no change to deposition in the distal conducting airways in the unconstricted lobes. Impaction is the dominant mechanism in the central airways, so deposition there is much more sensitive to increased flow rates. In fact, figure \ref{fig:PatchyCPA}(c) suggests that the increase in deposition in the central airways in these lobes is at least as large as the decrease in the lobes with constrictions. Within the constricted lobes, there are still some open paths, and deposition in airways on these paths still tends to increase as long as they are not directly proximal to a constriction (figure \ref{fig:PatchyCPA}d). Therefore, some inhaled particles continue to be deposited in a lobe even if many of its airways are constricted, although deposition then strongly favours the open airways that are not on the same path as any constriction. The lack of increase in distal-airways deposition in the lobes without constrictions (figure \ref{fig:PatchyCPA}c) is due to the fact that increased flow rates mean shorter residence times for most inhaled particles so decreased sedimentation, which is the dominant deposition mechanism in the distal airways. In turn, this means residence times in the acini are longer in these lobes, leading to increased acinar deposition. In these lobes, the ratio of central to distal conducting airway deposition is increased purely by application of constrictions elsewhere in the lungs.

\section{Discussion}\label{sec:discussion}

We have developed a model of inhaled particle deposition which predicts ventilation patterns based on the resistance of every airway in the lungs. {Existing computational models, such as the Multiple-Path Particle Dosimetry (MPPD) model \cite{asgharian2001particle}, have been built on assumptions that the lungs are healthy and that ventilation is uniform. Such models cannot predict the impacts of airway constriction and blockage on ventilation and particle deposition. Our model, therefore, provides a first step towards predicting particle deposition in patients with obstructive lung disease. The model outlined here provides a framework that can in future be extended to incorporate more clinically relevant physics, and to integrate clinical measurements of lung function and ventilation heterogeneity to move towards patient-specific predictions of drug deposition in obstructive lung diseases. }

We have tested the model in several lung geometries, validating total deposition results in healthy lung simulations against MPPD \cite{mppd_v3.04}, showing good agreement even despite using different lung geometries. We investigated the impact of applying various patterns of distal airway constrictions, representative of features of CF disease. Strong changes in deposition can occur in the constricted airways themselves, but also, when changes to ventilation are taken into account, there can be significant impacts on deposition in unconstricted airways elsewhere in the lungs. Deposition in airways directly proximal or distal to constricted airways is reduced as flow rates through these paths are decreased. The severity of constriction at which this decrease becomes severe approximately coincides with a spike in the predicted LCI. We demonstrated that when distal airway constrictions are clustered together, their effects on deposition in the central airways can be heightened, with deposition becoming more heterogeneous. Central airways away from the constrictions received significantly increased deposition, driven by enhanced inertial impaction as flow rates were increased through these open paths, since constricted airways received less flow and the flow rate into the trachea was assumed to be the same before and after applying constrictions. By contrast, the unconstricted distal conducting airways tended not to receive an increased dose since gravitational sedimentation, the dominant deposition mechanism in the small airways, is not increased by faster flow. {This disparity between the impact on central-airways and distal-airways deposition away from constrictions highlights a mechanism by which an increase in central-to-peripheral deposition ratio may be driven by the presence of localised severe distal airway constrictions, and our simulation predictions provide some quantification of this effect.}

Our results highlight the importance of accounting for altered ventilation patterns when simulating particle deposition in lungs affected by airway disease. Experimental studies have previously established that CF disease can significantly alter sites of particle deposition \cite{anderson1989effect}. We found that severe constrictions in the lungs' upper lobes, which are often affected early in CF \cite{turcios2020cystic}, can cause increased deposition in the central airways elsewhere in the lungs. This may partially explain the increased central-airways deposition observed in some severe CF patients \cite{anderson1989effect}. However, other effects, such as increased turbulence, which our model cannot account for, may also contribute. Further work directly replicating the experiments of Anderson et al. \cite{anderson1989effect} would be required to confirm this. In severe disease, breathing may become faster and shallower \cite{hart2002changes}, which may also enhance central-airways deposition. 

{Our model provides detailed predictions of deposition patterns throughout the central and distal airways, which have a significantly finer resolution than can be achieved with current medical imaging techniques.}
%Our model provides detailed predictions of deposition down to the single-airway scale, beyond the scope of current medical imaging techniques. %We highlighted some potential limitations of gamma scintigraphy by simulating it in lungs with clusters of airway constriction: smaller clusters of constrictions can have some impact on the distribution of particle deposition (figure \ref{fig:PatchyCPA}c), but they may not always be evident from scintigraphy (figure \ref{fig:PatchyCPA}a). 
The physical insights we have provided, relating changes in ventilation and deposition patterns, have the potential to improve estimates of regional dosing of an inhaled therapeutic once ventilation patterns have been inferred in a patient's lungs from clinical tests of ventilation heterogeneity, such as MBW or ventilation MRI \cite{horsley2009lung,smith2018patterns}. Dosing plans drawn up based on lung exposure calculated from MPPD and healthy volunteer studies may not be accurate for those with significant ventilation heterogeneity. Our data suggest there may be particular risks around higher concentrations of inhaled drug being delivered to central airways and lower doses to the most strongly affected lung regions. 

%\color{black}
Some relevant physical effects are not included in the current model, but could be incorporated in future: for example, airway wall elasticity \cite{venegas2005self}, which could enable modelling of transient airway collapse; mechanical interdependence of the acini, which has been incorporated in previous ventilation models \cite{roth2017comprehensive}; and extrathoracic deposition, which may be significant for large particles \cite{darquenne_2012_aerosol}. Incorporating extrathoracic deposition could enable direct comparison with experimental data in patients; here, we have compared deposition results against MPPD simulations without extrathoracic deposition, which still provides good validation since MPPD compares well with experimental data \cite{asgharian_2022_mixing,asgharian2006prediction}. To account for the effects of boundary-layer phenomena, plug-like flow, or the increase in resistance due to turbulent dissipation, particularly in the large airways, Poiseuille's law could be replaced by an alternative resistance law \cite{pedley1970prediction,reynolds1979modeling,lambert1982computational,ismail2013coupled}. The impacts of altered airway wall mechanics on particle deposition may be particularly relevant if extending the model to describe drug delivery in asthma \cite{venegas2005self}, and incorporating spatially varying elastic properties into the acinar model could enable modelling of heterogeneous emphysema in COPD \cite{burrowes2014computational}. {Whilst we have applied fixed patterns of airway constrictions in this study, incorporating a model for the formation of airway blockages, including modelling transient airway collapse and reopening, could be important for extending the model to capture a wider range of diseases and disease states in future.}

 We have provided detailed physical insight into the effects of simulated patterns of airway constriction, representative of typical features of CF. %In future, inferring patterns of constriction directly from medical imaging could allow us to move towards patient-specific predictions of the effects of disease. 
{In order to isolate the impacts of airway constriction, we have not varied parameters such as breathing rate or tidal volume. Faster, shallower breathing should be incorporated if modelling patients or patient-phenotypes where this is observed \cite{hart2002changes}; this study focused on features representative of mild disease, where breathing rates are unlikely to be significantly altered.} We have demonstrated that the model can simulate deposition of particles of a wide range of sizes (figure \ref{fig:Uncon}e), but have focused on assessing the impacts of airway constriction on deposition of $4\mu\mathrm{m}$-diameter particles. Systematic investigation of the effects of varying particle size in lung models with airway constrictions is left for a future study. By assessing the impacts of breathing rates and particle size in diseased lungs, modelling results could then be used to suggest new clinical strategies for drug delivery.

{The model presented here provides an advance on previous models, such as single-path \cite{yeh1980models} or multiple-path \cite{asgharian2001particle} models, by incorporating a physics-based ventilation model that accounts for the resistance of each individual airway. This means that it can predict the impacts of spatially heterogeneous patterns of airway constrictions or blockages on ventilation and particle deposition throughout the lungs. Models incorporating three-dimensional CFD simulations in the central airways provide more detailed predictions of deposition patterns, but the computational cost of CFD generally means only a small number of airways can be simulated \cite{longest2012silico}. Experimental and numerical studies have shown that turbulence in the upper airways can enhance rates of particle deposition \cite{darquenne2004aerosol}, and CFD models may be able to capture these effects. Models that couple CFD in the large airways with reduced-order models for the distal airways can simulate deposition throughout the lungs on inhalation and exhalation \cite[e.g., ][]{oakes_aerosol_2017,kuprat2021efficient}, but the whole-lung models of Oakes et al. \cite{oakes_aerosol_2017,poorbahrami2021whole} or Kuprat et al. \cite{kuprat2021efficient} assume uniform ventilation within the distal airways, so these models could not have responded to distal airway constrictions in a physics-based way, as our model does. Our reduced-order modelling approach lowers the computational time for our simulations compared to most CFD models, enabling us to run a larger number of simulations to investigate different patterns of airway constriction. %A new approach proposed by Grill et al. \cite{grill2023silico} has the potential to also incorporate ventilation heterogeneity, but has not been used to explore the impacts of small airway constriction or airway blockage. 
}

Using a novel computational model that accounts for altered ventilation induced by airway constrictions, we have demonstrated some of the key impacts of airway disease on patterns of inhaled particle deposition. We have demonstrated how airway constrictions localised to the distal airways can affect deposition throughout the lungs. Spatial clustering of constrictions can increase the impact on central-airways deposition, making it more heterogeneous by reducing deposition in airways directly proximal to constrictions and increasing deposition elsewhere. These results have implications for understanding how ventilation defects and airway blockage in obstructive lung diseases may affect how inhaled therapeutics deposit. This is an important step towards developing realistic simulations of particle and drug deposition that can be used to accurately model exposure in disease states as well as healthy lungs.

%\appendix

%\end{multicols}

\enlargethispage{20pt}

%\ethics{Insert ethics text here.}

\dataccess{The code used to simulate ventilation and particle transport and deposition is available via xxxxxx. Code used to generate airway constriction patterns, the network files used in all simulations and the code used to generate simulated scintigraphy plots are available via xxxxxx. Generation of the lung networks used the same method as Whitfield et al. \cite{whitfield2020spectral}; the code used to generate those networks was published previously here https://dx.doi.org/10.5281/zenodo.3709073, with the airway centre-line data from CT and computed networks here https://dx.doi.org/10.5281/zenodo.3709105.}

\aucontribute{J.D.S. led the investigation, wrote the paper, and contributed to the design of the study and coding of the numerical algorithms. A.H. critically revised the manuscript and contributed to the development of the study. O.E.J. contributed to the conception, design and coordination of the study, and critically revised the manuscript. A.B.T. contributed to the design and coordination of the study and critically revised the manuscript. J.M.W. collected and collated the CT data used in the study. C.A.W. carried out the CT image processing, and contributed to the conception, design and coordination of the study, the coding of the numerical algorithms, and revision of the manuscript. }

\competing{We declare we have no competing interests.}

\funding{J.D.S. was supported by an EPSRC doctoral training award. C.A.W. is funded by, and A.H. and O.E.J. are supported by, the National Institute for Health and Care Research (NIHR) Manchester Biomedical Research Centre (BRC) (NIHR203308). {A.H. and C.A.W. are supported by the ESPRC Network Plus ``BIOREME'' (EP/W000490/1). A.H. and C.A.W. also acknowledge funding from the Cystic Fibrosis Trust and Cystic Fibrosis Foundation through the Strategic Research Centre grant SRC025, which is supporting the continued development and translation of the research presented in this paper.} The numerical algorithms developed extend on previous work by C.A.W. funded by the UK Medical Research Council (MR/R024944/1). }

%\ack{Insert acknowledgment text here.}

\disclaimer{The views expressed are those of the authors and not necessarily those of the NIHR or the Department of Health and Social Care.}

%%%%%%%%%% Insert bibliography here %%%%%%%%%%%%%%

%\bibliography{refs}
%\bibliographystyle{RS}
\printbibliography

\end{document}

\maketitle	
%\pagebreak

% Optional TOC
% \tableofcontents
% \pagebreak

%--Paper--

%\section{Model setup}

\section{Fundamentals of calculus on discrete networks}
\label{sec:discrete_calc}

We model the lungs as a discrete network of edges and vertices. In this section, we introduce some basic concepts of calculus on discrete networks, which we utilise below when detailing the model equations. 

Consider a network $\mathcal{N}=(\mathcal{E},\mathcal{V})$, composed of a set of edges, $\mathcal{E}$, and a set of vertices, $\mathcal{V}$. Each edge, $e_j\in\mathcal{E}$, is connected to exactly two vertices, $e_j=\{v_k,v_{k'}\}$, and the edge has an orientation such that $v_k$ is at the proximal end relative to vertex $v_1$ while $v_{k'}$ is at the distal end relative to $v_1$. The vertex $v_1$ is the inlet vertex, which represents the top of the trachea. The topology of the network is described by the incidence matrix, $\mathsf{N}\in\mathbb{R}^{|\mathcal{E}|\times|\mathcal{V}|}$, which has components,
\begin{align}
    {{N}}_{jk} = \begin{cases}
    1 \quad &\mbox{if} \; {e}_j = \{{v}_k,{v}_{k'}\}\,\,\mbox{for some}\,\, v_{k'}\in\mathcal{V},\\
    -1 \quad &\mathrm{if} \; {e}_j = \{{v}_{k'},{v}_k\}\,\,\mbox{for some}\,\, v_{k'}\in\mathcal{V},\\
    0 \quad &\mathrm{if} \; {v}_k \notin {e}_j.
    \end{cases}
    \label{incidence}
\end{align}
Each row of $\mathsf{N}$ corresponds to an edge in the network, and contains exactly one entry of $-1$, at the location corresponding to the vertex at the distal end of the edge, and exactly one entry of $1$, at the location corresponding to the vertex at the proximal end of the edge. We additionally define the unsigned incidence matrix, $\mathsf{N}_{\rm u}$, via ${N}_{\mathrm{u},jk}=|{N}_{jk}|$. 

\begin{figure}
    \centering
    \includegraphics[width = \textwidth]{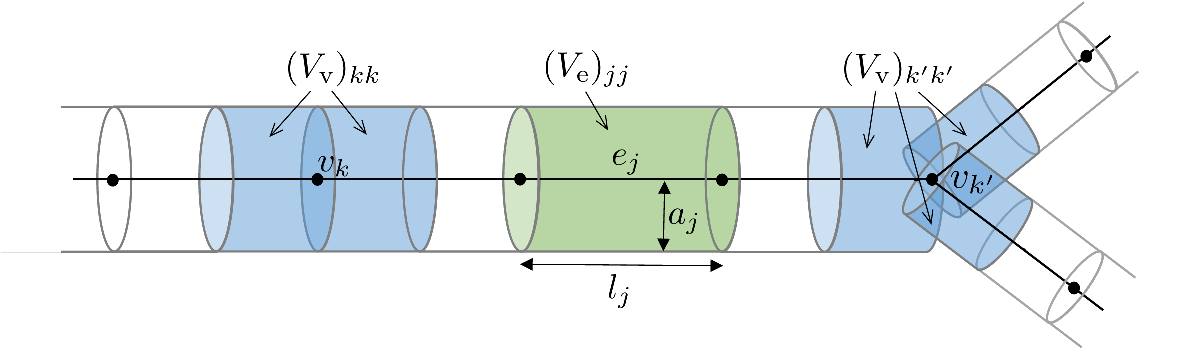}
    \caption{Sketch illustrating how volumes are assigned to the edges and vertices of the network. The cylindrical volume associated with an edge $e_j$ is highlighted in green. The volumes associated with two vertices, labelled $v_k$ and $v_{k’}$ are highlighted in blue: these volumes consist of the sum of half of the volumes of each edge connected to the given vertex. }
    \label{fig:VolSketch}
\end{figure}

Each edge, $e_j\in\mathcal{E}$, has a fixed length ${l}_j$ and radius ${a}_j$ and, hence, a cross-sectional area of $ \pi {a}_j^2$ and a volume of $\pi {l}_j{a}_j^2$. We define the discrete scalar fields, $\mathsf{l},  \mathsf{a}\in\mathbb{R}^{|\mathcal{E}|}$, containing the lengths and radii of each edge, and we define two diagonal matrices, $\mathsf{S},\mathsf{V}_{\rm e}\in\mathbb{R}^{|\mathcal{E}|\times|\mathcal{E}|}$, with components equal to the edge cross-sectional areas and volumes, respectively,
\begin{equation}
    S_{jj} = \pi {a}_j^2 \quad\mbox{and}\quad \left({V}_{\rm e}\right)_{jj} = \pi {l}_j{a}_j^2.
\end{equation}
We also define a diagonal matrix, $\mathsf{V}_{\rm v}\in\mathbb{R}^{|\mathcal{V}|\times|\mathcal{V}|}$ of \say{vertex volumes}, as
\begin{equation}
    \mathsf{V}_{\rm v} \equiv \mathrm{diag}\left[\frac{1}{2}\mathsf{N}_{\rm u}^T\mathsf{V}_{\rm e}\mathbbm{1}_{\rm e}\right],
    \label{Vvdefn}
\end{equation}
where $\mathbbm{1}_{\rm e} \in\mathbb{R}^{|\mathcal{E}|}$ is the vector of ones, $\mathbbm{1}_{\rm e}=(1,1,\dots,1)$. The component $\left(V_{\rm v}\right)_{kk}$ is equal to the sum of half of the volumes of the edges connected to vertex $v_k$. Figure 
\ref{fig:VolSketch} illustrates how this volume is assigned to a vertex in the case that the vertex is in the interior of an airway or at the junction between airways. %We define two diagonal matrices,
%\begin{equation}
%    \mathsf{V}_{\mathrm{v}} = \mathrm{diag}\left(\mathsf{v}_{\mathrm{v}}\right)\quad\mbox{and}\quad \mathsf{V}_{\mathrm{e}} = \mathrm{diag}\left(\mathsf{v}_{\mathrm{e}}\right),
%    \label{VvVedefns}
%\end{equation}
%which are vertex-volume and edge-volume metric tensors, respectively, which we will now use to define inner products and operators on the network. 

We define two inner products on the network in order to construct mimetic finite differences \cite{lipnikov2014mimetic}. First, given two discrete scalar fields defined on the vertices, $\mathsf{p},\mathsf{q}\in\mathbb{R}^{|\mathcal{V}|}$, we define their inner product as
\begin{equation}
    \langle\mathsf{p},\mathsf{q}\rangle_{\rm v} \equiv \mathsf{p}^T\mathsf{V}_{\rm v}\mathsf{q} = \sum_{k=1}^{|\mathcal{V}|} {p}_k\left({V}_{\rm v}\right)_{kk}{q}_k.
    \label{innerprodv}
\end{equation}
Similarly, given two fields defined on the edges, $\mathsf{f},\mathsf{g}\in\mathbb{R}^{|\mathcal{E}|}$, we define their inner product as
\begin{equation}
    \langle\mathsf{f},\mathsf{g}\rangle_{\rm e} \equiv \mathsf{f}^T\mathsf{V}_{\rm e}\mathsf{g} = \sum_{j=1}^{|\mathcal{E}|} {f}_j\left({V}_{\rm e}\right)_{jj}{g}_j.
    \label{innerprode}
\end{equation}

The incidence matrix, $\mathsf{N}$, acts as a difference operator on a vertex vector, mapping a quantity defined on the vertices to the difference of that quantity across each of the edges. We can, thus, construct a gradient operator,
\begin{equation}
    \mathrm{grad}(\mathsf{p}) \equiv\mathrm{diag}\left(\mathsf{l}\right)^{-1}\mathsf{N}\mathsf{p},
    \label{gradOperator}
\end{equation}
given any $\mathsf{p}\in\mathbb{R}^{|\mathcal{V}|}$. The operator \eqref{gradOperator} is a discrete analogue of the continuous gradient operator that acts on a continuous field. For example, suppose some continuous quantity, $p(z,t)$, is defined in an airway, with $z$ being the coordinate pointing along the axis of the airway. If we discretise $p(z,t)$ by defining a vector $\mathsf{p}(t)\in\mathbb{R}^{|\mathcal{V}|}$ of its values at several vertices evenly spaced along the airway, then \eqref{gradOperator} provides a central difference approximation, centred at the mid-point of each edge of the airway, to $\partial_zp$. 

Using \eqref{innerprodv} and \eqref{innerprode}, we also define a discrete divergence operator,
\begin{equation}
    \mathrm{div}(\mathsf{f}) \equiv -\mathsf{V}_{\rm v}^{-1}\mathsf{N}^T\mathrm{diag}(\mathsf{s})\mathsf{f},
    \label{DivOperator}
\end{equation}
for any vector $\mathsf{f}\in\mathbb{R}^{|\mathcal{E}|}$. The divergence operator \eqref{DivOperator} is related to the gradient operator \eqref{gradOperator} via
\begin{equation}
    \langle \mathsf{f},\mathrm{grad}(\mathsf{p})\rangle_{\rm e} = \langle -\mathrm{div}(\mathsf{f}),\mathsf{p}\rangle_{\rm v},
\end{equation}
for any vectors $\mathsf{f}\in\mathbb{R}^{|\mathcal{E}|}$ and $\mathsf{p}\in\mathbb{R}^{|\mathcal{V}|}$. The operator \eqref{DivOperator} is the discrete analogue of the continuous divergence operator. To see this, again consider a single airway, discretised into a number of edges and vertices, and suppose we have a continuous function, $f(z,t)$, defined along the length of the airway. If $f$ is, say, a flux of some quantity through the airway, it is natural to discretise it by taking its value at the mid-point of each edge and defining an edge vector, $\mathsf{f}(t)\in\mathbb{R}^{|\mathcal{E}|}$, with entries equal to the flux at the edge mid-points. Since the edges within the airway all have the same length, each vertex in the interior of the airway has the same volume as one of the edges. Therefore, $\mathrm{div}(\mathsf{f})$ provides a second-order finite difference approximation, centred at the vertices in the airway, to the derivative $\partial_zf$. The operator \eqref{DivOperator} can also be used to mimic the divergence theorem. Suppose we have a set of adjacent vertices within the network, and a vector $\mathsf{d}\in\mathbb{R}^{|\mathcal{V}|}$ that has an entry $1$ corresponding to each of those vertices and a zero for every other vertex in the network. If $\mathsf{f}\in\mathbb{R}^{|\mathcal{E}|}$ is some flux defined on the network's edges, then $(\mathsf{Nd})^T\mathsf{f}$ is the net flux through the boundaries of the set of vertices represented by $\mathsf{d}$, and
\begin{equation}
    \mathsf{d}^T\mathsf{V}_{\rm v}\mathrm{div}(\mathsf{f}) = - (\mathsf{Nd})^T\mathsf{f},\label{Divthmanalgoue}
\end{equation}
is an analogous to a statement of the divergence theorem. 
%With \eqref{DivOperator}, we can also state a discrete version of the divergence theorem as follows. Given a set of edges, $\hat{\mathcal{E}}$, which has a set of vertices, $\hat{\mathcal{S}}$, as its boundary. Then, from the definition \eqref{DivOperator},

  %At the vertices corresponding to the end-points of the airway, where each vertex is connected to more than two edges, a simple correspondence with a finite difference scheme is not possible. However, we will see below that when a discrete transport equation is written in conservative form, the $\mathrm{div}(\cdot)$ operator acts on the flux to conserve mass exactly through the airway end-point vertices as well as through the interior vertices. 

\begin{figure}
    \centering
    \includegraphics[width = \textwidth]{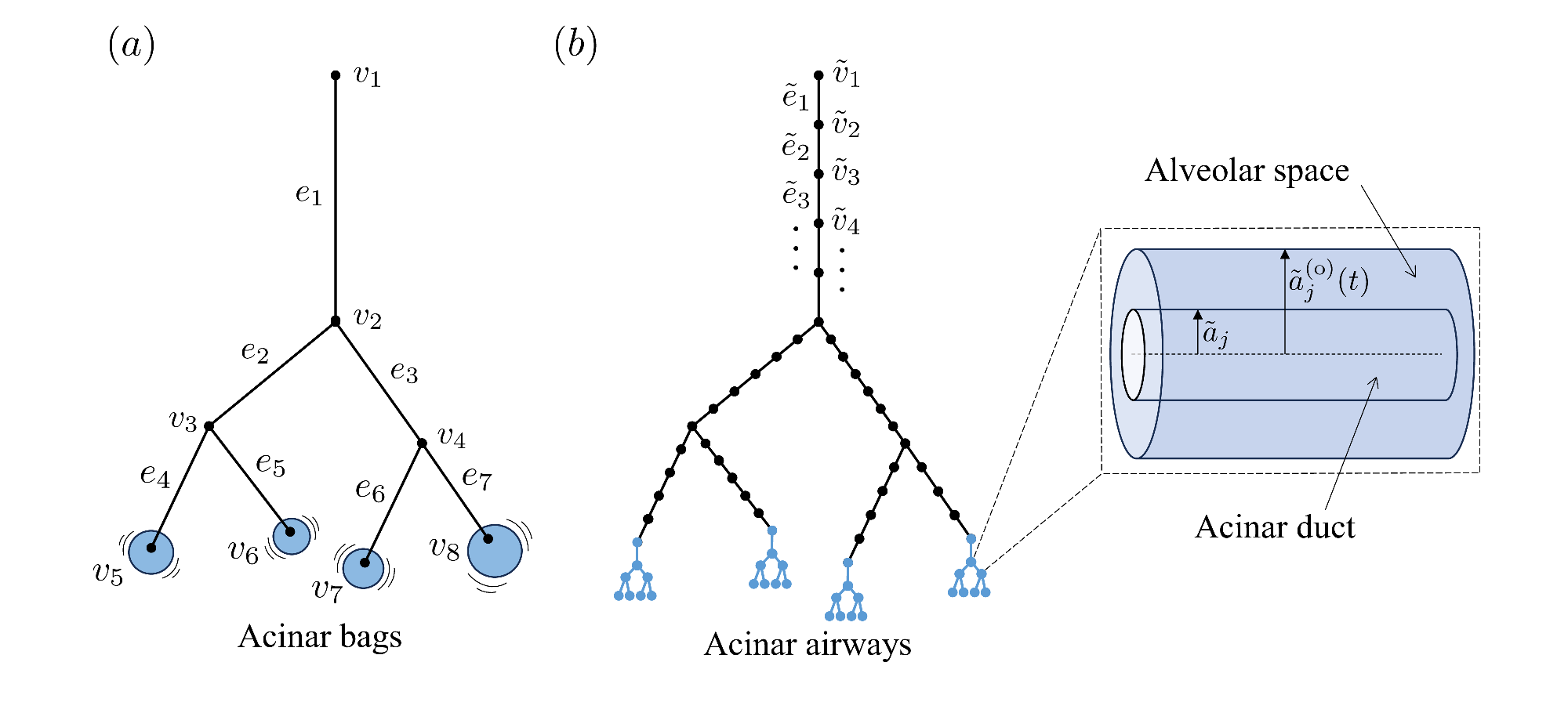}
    \caption{Sketches of the lung network geometry. The conducting airway network is plotted in black, with vertices represented by dots and edges by line segments. (a) The network used to simulate ventilation, with each terminal vertex representing a single acinus, and the acini modelled as elastic bags that can expand and contract. (b) The modified network used to simulate particle transport and deposition, with the acini modelled as symmetric airway trees. The inset in (b) illustrates the structure of an acinar airway: a duct with a fixed radius, $\tilde{a}_j$, and a time-dependent outer radius, $\tilde{a}_j^{(\rm o)}(t)$, where the space between the duct radius and outer radius represents alveolar space. }
    \label{fig:networks}
\end{figure}

\section{Ventilation model}
\label{sec:vent_equations}

\subsection{Model setup}

The model of ventilation is substantially the same as that of Whitfield et al. \cite{whitfield2020spectral} but we outline it here for completeness. The network on which we solve the ventilation model equations treats each airway as a single edge of the network, and each of the terminal vertices of the network represents a single acinus. In §\ref{sec:transport} below, we modify this network slightly to accurately simulate particle transport and deposition calculation. In the ventilation model, the acini are modelled as viscoelastic bags that expand and contract independently. Figure \ref{fig:networks}(a) illustrates the ventilation network geometry. We call this network $\mathcal{N} = (\mathcal{E},\mathcal{V})$, where $\mathcal{E}$ is the set of edges comprising the conducting airways, and $\mathcal{V}=\{v_1,\mathcal{V}_{\rm int},\mathcal{V}_{\rm term}\}$ is the set of vertices, with $v_1$ being the inlet vertex representing the top of the trachea, $\mathcal{V}_{\rm int}$ being the interior vertices representing bifurcations between airways, and $\mathcal{V}_{\rm term}$ being the terminal vertices representing acini. 

We assume a linear relationship between the pressure drop across, and the flux of air through, the conducting airways,
\begin{equation}
    {\mathsf{N}}\mathsf{p} = \mathrm{diag}(\mathsf{r})\mathsf{q}_{\rm e},
    \label{Prq1}
\end{equation}
where $\mathsf{N}$ is the incidence matrix for the network $\mathcal{N}$, $\mathsf{p}(t)\in\mathbb{R}^{|\mathcal{V}|}$ is the vector of pressures at the vertices, $\mathsf{q}_{\rm e}(t)\in\mathbb{R}^{|\mathcal{E}|}$ is the vector of volume fluxes through the edges, and $\mathsf{r}\in\mathbb{R}^{|\mathcal{E}|}$ is the vector of edge resistances. We approximate the resistance of an edge using Poiseuille's law, so ${r}_j = {8\mu_{\rm a} {l}_j}/{(\pi {a}_j^4)}$, where $\mu_{\rm a}$ is the air viscosity. %This simple assumption minimises the computational time of the simulations while still allowing the model to capture increased airway resistances caused by airway constrictions. 

We define the vector, $\mathsf{q}_{\rm v}(t) \equiv \mathsf{N}^T\mathsf{q}_{\rm e}\in\mathbb{R}^{|\mathcal{V}|}$, of net fluxes into each of the vertices. Then, we write the vertex flux vector in block form,
\begin{equation}
    %\mathsf{p} = 
    %\begin{pmatrix}
    %\mathsf{p}_0\\
    %\mathsf{p}_{\rm int}\\
    %\mathsf{p}_{\rm term}
    %\end{pmatrix},
    %\quad
    \mathsf{q}_{\rm v} = 
    \begin{pmatrix}
    {q}_1\\
    \mathsf{q}_{\rm int}\\
    \mathsf{q}_{\rm term}
    \end{pmatrix},
    \label{qvblockdefn}
\end{equation}
where ${q}_1(t)$ is the net flux into vertex $v_1$, $\mathsf{q}_{\rm int}(t)$ is the vector of net fluxes into the interior vertices $\mathcal{V}_{\rm int}$, and $\mathsf{q}_{\rm term}(t)$ is the vector of fluxes into the acini. We enforce mass conservation throughout the network via
\begin{equation}
    {q}_1 = Q_1(t),\quad \mathsf{q}_{\rm int} = \mathsf{0},\quad \mathsf{q}_{\rm term} = \frac{\mathrm{d} \mathsf{v}_{\rm term}}{\mathrm{d}t},
    \label{qmasscons}
\end{equation}
where $Q_1(t)$ is the flow rate at the top of the trachea and $\mathsf{v}_{\rm term}(t)\in\mathbb{R}^{|\mathcal{V}_{\rm term}|}$ is the vector of volumes of each of the acini. Note that $\mathbbm{1}_{\rm v}^T\mathsf{N}^T = \mathsf{0}$, from the definition of the incidence matrix, so $\mathbbm{1}_{\rm v}^T\mathsf{q}_{\rm v} = \mathsf{0}$, which states that global mass conservation is satisfied, with any volume of air that enters via the inlet vertex balanced by the change in volume of the terminal vertices.

It remains to state the relationship between the pressure in the acini and their volumes. Similarly to \eqref{qvblockdefn}, we write the vector of pressures in block form,
\begin{equation}
    \mathsf{p} = 
    \begin{pmatrix}
    {p}_1\\
    \mathsf{p}_{\rm int}\\
    \mathsf{p}_{\rm term}
    \end{pmatrix},
\end{equation}
where ${p}_1$ is the pressure at vertex $v_1$, $\mathsf{p}_{\rm int}(t)$ is the vector of pressures at the interior vertices $\mathcal{V}_{\rm int}$, and $\mathsf{p}_{\rm term}(t)$ is the vector of pressures in the acini. We assume that the acini expand and contract independently of each other, and that their expansion can be modelled using linear elasticity via
\begin{equation}
    R\frac{\mathrm{d}\mathsf{v}_{\rm term}}{\mathrm{d}t} + K\mathsf{v}_{\rm term} = \mathsf{p}_{\rm term} - \mathbbm{1}_{\rm term}P_{\rm pl}(t),
\end{equation}
where $R$ and $K$ are the resistance and elastance, respectively, of each acinus, assumed to be the same for all acini, $P_{\rm pl}(t)$ is the pleural pressure, also assumed to be uniform across all of the acini, and $\mathbbm{1}_{\rm term}\in\mathbb{R}^{|\mathcal{V}_{\rm term}|}$ is a vector of ones, $\mathbbm{1}_{\rm term}=(1,1,\dots,1)$. Finally, note that the pressure at the inlet vertex, ${p}_1$, is an arbitrary constant, so we set ${p}_1=0$.

Applying the operator $\mathsf{N}^T\mathrm{diag}(\mathsf{r})^{-1}$ to equation \eqref{Prq1} gives
\begin{equation}
    \mathsf{L}\mathsf{p} = \mathsf{q}_{\rm v},
    \label{Lpqv}
\end{equation}
where $\mathsf{L} = \mathsf{N}^T\mathrm{diag}(\mathsf{r})^{-1}\mathsf{N}$, so \eqref{Lpqv} is an equivalent statement to \eqref{Prq1} but relates the pressures at the vertices to the net fluxes into the vertices. When coupled with \eqref{qmasscons}, the relation \eqref{Lpqv} automatically enforces mass conservation at every vertex. Now, we can write the whole system of equations \eqref{qvblockdefn}-\eqref{Lpqv} as 
\begin{equation}
\label{VentLinSys}
\begingroup
\renewcommand*{\arraystretch}{1.3}
    \left(\begin{array}{c c c|c c c}
    %\multicolumn{3}{c}{\multirow{3}{*}
    & & & \mathsf{0} & 0 \\
    & \mathsf{L} &  & \mathsf{0} & \mathsf{0}\\
    & & & -{\mathsf{I}}_{\rm term}\frac{\mathrm{d}}{\mathrm{d}t} & \mathsf{0}\\
    \hline
    \mathsf{0}\quad & \mathsf{0} & -{\mathsf{I}}_{\rm term} & {\mathsf{I}}_{\rm term}({K}+{R}\frac{\mathrm{d}}{\mathrm{d}t}) & {\boldsymbol{\mathbbm{1}}}_{\rm term}\\
    %0 & 0 & 0 & \tilde{\mathsf{I}}_{\rm term} & \tilde{\mathsf{I}}_{\rm term}\frac{\mathrm{d}}{\mathrm{d}t} & 0\\
    1 \quad& \mathsf{0} & \mathsf{0} & \mathsf{0} & 0 \\
    \end{array}\right)
    \begin{pmatrix}
        {p}_1\\
        {\mathsf{p}}_{\rm int} \\
        {\mathsf{p}}_{\rm term} \\
        %\tilde{\boldsymbol{Q}}_{\rm term}\\
        \mathsf{v}_{\rm term}\\
        P_{\rm pl}\\
    \end{pmatrix}
    =
    \begin{pmatrix}
        {Q}_{1}\\
        \mathsf{0}\\
        \mathsf{0}\\
        \mathsf{0}\\
        0\\
    \end{pmatrix},
    \endgroup
\end{equation}
where ${\mathsf{I}}_{\rm term}\in\mathbb{R}^{|{\mathcal{V}}_{\rm term}|\times|{\mathcal{V}}_{\rm term}|}$ is the identity matrix. %To close the ventilation problem \eqref{VentLinSys} we must either impose the flow rate at the trachea, $Q_0(t)$, or the driving pleural pressure, $P_{\rm pl}(t)$. Unless otherwise stated, we impose the inlet flow rate to be
Finally, we close the system \eqref{VentLinSys} by imposing the flow rate into the trachea,
\begin{equation}
    Q_1(t) = \frac{\pi V_T}{T_b}\sin\left(\frac{2\pi t}{T_b}\right),\label{FlowBC}
\end{equation}
where $V_T$ is the tidal volume and $T_b$ is the breath time. 

\subsection{Solution methods}

To solve \eqref{VentLinSys}-\eqref{FlowBC}, we follow the approach outlined by Whitfield et al. \cite{whitfield2020spectral}, assuming that $t\gg T_b$ so that the solution is periodic in time, with period $T_b$. We look for solutions of the form
\begin{eqnarray}
    \mathsf{p}_{\rm int} = \mathsf{p}_{\rm int}^{(0)} + \mathsf{p}_{\rm int}^{(c)}\cos{\left(\frac{2\pi t}{T_b}\right)}
    +\mathsf{p}_{\rm int}^{(s)}\sin{\left(\frac{2\pi t}{T_b}\right)},\label{periodic_pint}\\
    \mathsf{p}_{\rm term} = \mathsf{p}_{\rm term}^{(0)} + \mathsf{p}_{\rm term}^{(c)}\cos{\left(\frac{2\pi t}{T_b}\right)}
    +\mathsf{p}_{\rm term}^{(s)}\sin{\left(\frac{2\pi t}{T_b}\right)},\\
    \mathsf{v}_{\rm term} = \mathsf{v}_{\rm term}^{(0)} + \mathsf{v}_{\rm term}^{(c)}\cos{\left(\frac{2\pi t}{T_b}\right)}
    +\mathsf{v}_{\rm term}^{(s)}\sin{\left(\frac{2\pi t}{T_b}\right)},\\
    P_{\rm pl} = P_{\rm pl}^{(0)} + P_{\rm pl}^{(c)}\cos{\left(\frac{2\pi t}{T_b}\right)}
    +P_{\rm pl}^{(s)}\sin{\left(\frac{2\pi t}{T_b}\right)},\label{periodic_Ppl}
\end{eqnarray}
where the coefficients $\{\mathsf{p}_{\rm int}^{(0)},\mathsf{p}_{\rm int}^{(c)},\mathsf{p}_{\rm int}^{(s)},\mathsf{p}_{\rm term}^{(0)},\mathsf{p}_{\rm term}^{(c)},\mathsf{p}_{\rm term}^{(s)},\mathsf{v}_{\rm term}^{(0)},\mathsf{v}_{\rm term}^{(c)},\mathsf{v}_{\rm term}^{(s)}\}$ are all constant vectors, and $P_{\rm pl}^{(0)}, P_{\rm pl}^{(c)}$ and $P_{\rm pl}^{(s)}$ are also constant coefficients. Substituting \eqref{periodic_pint}-\eqref{periodic_Ppl} into \eqref{VentLinSys}, gives a linear system of equations for all of these constant coefficients, which we solve using the SparseLU solver in the C++ library Eigen \cite{eigen_2010_v3}. In all simulations, we take $T_b=5\mathrm{s}$, $V_T=0.625\mathrm{mL}$, $R_{\rm tot}=0.6\mathrm{cmH}_2\mathrm{OsL}^{-1}$, where $R_{\rm tot}$ is the total resistance of all of the acini combined, and $K_{\rm tot}=6.82\mathrm{cmH}_2\mathrm{OL}^{-1}$, where $K_{\rm tot}$ is the total combined elastance of all of the acini, and each acinus has the same individual resistance and elastance.

\section{Particle transport and deposition model}\label{sec:transport}

\subsection{Model setup}

After solving the ventilation problem \eqref{VentLinSys}, we use the computed flow rates in each airway to solve for the transport of an inhaled gas, and deposition of particles from that inhaled gas. In this section, we will present the equations describing the transport and deposition of inhaled particles. When solving the transport equations we no longer model the acini as single vertices, as we did above. Instead, the acini are modelled as symmetric networks of airways with alveolar space attached to each acinar airway (see sketch in figure \ref{fig:networks}b). Each acinus has the same structure, with eight generations of airways, and the lengths $\{\tilde{l}_j\}$ and radii $\{\tilde{a}_j\}$ of the acinar airways in each generation are taken from \cite{haefeli1988morphometry}. Each acinar airway is also assigned an \say{outer radius}, $\tilde{a}_j^{(\rm o)}(t)$, with the space between the airway radius and the outer radius representing alveolar space. The values of $\{\tilde{a}_j^{(\rm o)}(t)\}$ are determined by distributing the volumes of the acini, $\mathsf{v}_{\rm term}(t)$, from the solution of the ventilation problem \eqref{VentLinSys}, between the airways in each acinus in the modified network. The proportion of the acinus volume assigned to each generation of the acinar tree is also the same across all acini, and based on data from \cite{haefeli1988morphometry}. We denote this modified lung network by $\tilde{\mathcal{N}} = (\tilde{\mathcal{E}},\tilde{\mathcal{V}})$, where $\tilde{\mathcal{E}} = \{\mathcal{E},\mathcal{E}_{\rm acin}\}$ and $\tilde{\mathcal{V}} = \{v_1,\mathcal{V}_{\rm int},\mathcal{V}_{\rm acin}\}$, with $\mathcal{V}_{\rm acin}$ and $\mathcal{E}_{\rm acin}$ being the vertices and edges constituting the acinar airways. Quantities defined on the network $\tilde{\mathcal{N}}$ will typically be furnished with tildes to distinguish them from quantities previously defined on $\mathcal{N}$ in §\ref{sec:vent_equations}. 

We derive an Eulerian model for particle transport through the lungs, in which the concentration of inhaled particles is tracked at each point in the lungs. In a straight, cylindrical tube with an air flow velocity profile, $\boldsymbol{u}(\boldsymbol{x},t)$, the evolution of the concentration, $c(\boldsymbol{x},t)$, of particles is governed by
\begin{equation}
    \frac{\partial c}{\partial t} = \nabla\cdot\left( -\boldsymbol{u} c + D\nabla c\right),\quad \nabla\cdot\boldsymbol{u}=0
    \label{conttransporteqn}
\end{equation}
where 
\begin{equation}
    D = \frac{k_{\rm B}T\mu_{\rm a}C_{\rm c}}{3\pi d_{\rm p}}
    \label{particle_diffusivity}
\end{equation}
is the diffusivity of the particles, with $k_{\rm B}$ being the Boltzmann constant, $T$ the body temperature, $\mu_{\rm a}$ the viscosity of air, $d_{\rm p}$ the particle diameter and 
\begin{equation}
    C_{\rm c} = 1 + 2Kn\left(1.257 + 0.4e^{-0.55/Kn}\right)
\end{equation}
the Cunningham slip correction factor \cite{hinds2022aerosol} which takes account of non-continuum effects when the Knudsen number, $Kn = l_{\rm m}/d_{\rm p}$, where $l_{\rm m}$ is the mean free path, is small. Supposing the mean axial velocity is $\bar{u}(r,t)\boldsymbol{\hat{z}}$, \eqref{conttransporteqn} can be approximated by an effectively one dimensional transport equation,
\begin{equation}
    \frac{\partial \bar{c}}{\partial t} = \frac{\partial}{\partial z}\left(-\bar{u}\bar{c} + D_{\rm eff}\frac{\partial \bar{c}}{\partial z}\right) - s,
    \label{averageadvdiff}
\end{equation}
where the cross-sectional average of the concentration is
\begin{equation}
    \bar{c} = \frac{1}{\pi a^2}\int_0^ac\,r\mathrm{d}r,
\end{equation}
$a$ is the tube radius, and $D_{\rm eff}$ is an effective diffusivity. In the conducting airways, we approximate the effective diffusivity by empirical formulae due to Lee et al. \cite{lee2000dispersion,lee2001dispersion}, which are modifications of the formula of Scherer et al. \cite{scherer_1975_diffusivities}, that take into account dispersion of particles,
\begin{equation}
    D_{\rm eff} = \left\{\begin{array}{c}
        D + 0.7\bar{u}a \quad\mbox{if}\quad \bar{u} \geq 0,\\ 
        D + 0.26\bar{u}a \quad\mbox{if}\quad \bar{u} < 0,
    \end{array}\right.
    \label{Deff}
\end{equation}
where $\bar{u}>0$ corresponds to flow towards the distal end of the airway and $\bar{u}<0$ corresponds to flow towards the proximal end of the airway. In acinar airways, we instead set $D_{\rm eff} = D_{\rm a} = D + \sigma\bar{u}l/2$, using a formula derived by Taulbee \& Yu \cite{taulbee1975theory} for particle dispersion in symmetric trees, where $l$ is the airway length and $\sigma$ is a measure for the variability in airway radii and lengths, which is taken to be $\sigma=0.6$. Lastly in \eqref{averageadvdiff}, $s$ is a loss term describing particle deposition on the tube walls, which is defined below in §\ref{sec:deposition}. 

The discrete calculus machinery introduced in §\ref{sec:discrete_calc} is then used to provide a discretisation of the continuous advection-diffusion equation \eqref{averageadvdiff} in every airway of the network. We define concentrations on the vertices of the network, $\tilde{\mathsf{c}}(t)\in\mathbb{R}^{|\tilde{\mathcal{V}}|}$, and mean velocities, $\tilde{\mathsf{u}}(t)\in\mathbb{R}^{|\tilde{\mathcal{E}}|}$ on the edges. Then utilising the gradient and divergence operators \eqref{gradOperator} and \eqref{DivOperator}, the discrete analogue of the transport equation \eqref{conttransporteqn} can be written as
\begin{equation}
    \frac{\mathrm{d}\tilde{\mathsf{c}}}{\mathrm{d}t} = -\tilde{\mathsf{V}}_{\rm v}^{-1}\tilde{\mathsf{N}}^T\mathrm{diag}(\tilde{\mathsf{s}})\left[-\mathrm{diag}(\tilde{\mathsf{u}})\tilde{\mathsf{N}}_{\rm v}\tilde{\mathsf{c}} + \mathsf{D}_{\rm eff}\mathrm{diag}(\,\tilde{\mathsf{l}}\,)^{-1}\tilde{\mathsf{N}}\tilde{\mathsf{c}}\right] - \tilde{\mathsf{\Lambda}},
    \label{discretetransporteqn}
\end{equation}
where $\tilde{\mathsf{V}}_{\rm v}\in\mathbb{R}^{|\tilde{\mathcal{V}}|\times|\tilde{\mathcal{V}}|}$ is the diagonal vertex volume tensor as defined in \eqref{Vvdefn}, $\tilde{\mathsf{N}}\in\mathbb{R}^{|\tilde{\mathcal{E}}|\times|\tilde{\mathcal{V}}|}$ is the incidence matrix for the network $\tilde{\mathcal{N}}$ defined as in \eqref{incidence}, $\tilde{\mathsf{N}}_{\rm v}(t)\in\mathbb{R}^{|\tilde{\mathcal{E}}|\times|\tilde{\mathcal{V}}|}$ is a modified incidence matrix defined via 
\begin{equation}
    \tilde{{N}}_{\mathrm{v},jk} = \max\left[0,\mathrm{sgn}(\tilde{{u}}_j)\tilde{{N}}_{jk}\right],
    \label{Nvdefn}
\end{equation}
$\tilde{\mathsf{s}}\in\mathbb{R}^{|\tilde{\mathcal{E}}|}$ is the vector of edge cross-sectional areas, $\mathsf{D}_{\rm eff}(t)\in\mathbb{R}^{|\tilde{\mathcal{E}}|\times|\tilde{\mathcal{E}}|}$ is a diagonal matrix of effective diffusivities on each edge, defined by
\begin{equation}
    {D}_{\mathrm{eff},jj} = \left\{\begin{array}{c}
        D + 0.7\tilde{u}_j\tilde{a}_j \quad\mbox{if}\quad \tilde{u}_j \geq 0,\\ 
        D + 0.26\tilde{u}_j\tilde{a}_j \quad\mbox{if}\quad \tilde{u}_j < 0,
    \end{array}\right.
    \label{Deffdiscretedefn}
\end{equation}
if $\tilde{e}_j$ is a conducting airway, and $D_{\rm eff}=D + \sigma \tilde{l}_{\mathrm{a},j}\tilde{u}_j/2$ if $\tilde{e}_j$ is an acinar airway, where $\tilde{l}_{\mathrm{a},j}$ is the length of the airway containing $\tilde{e}_j$,
$\tilde{\mathsf{l}}\in\mathbb{R}^{|\tilde{\mathcal{E}}|}$ is the vector of edge lengths and $\tilde{\mathsf{\Lambda}}\in\mathbb{R}^{|\tilde{\mathcal{V}}|\times|\tilde{\mathcal{V}}|}$ is a diagonal matrix of loss terms at each vertex, which we will specify below. The term within the square brackets in \eqref{discretetransporteqn} represents the flux per unit area of concentration through each edge of the network, with the first term an advective flux and the second term a diffusive flux. In the advective flux term, the mean air velocities $\tilde{\mathsf{u}}(t)$ are calculated by taking the solution of the ventilation problem \eqref{VentLinSys}, calculating the flux of air through each edge using \eqref{Prq1} and then dividing those fluxes by the cross-sectional area of each edge. Velocities in the acinar airways, which are not directly calculated in the ventilation problem, are calculated at each time point via mass conservation, given the flux of air into each acinus and the volume of alveolar space in each airway, inferred from the outer radius $\tilde{a}^{(o)}(t)$. The modified incidence matrix \eqref{Nvdefn} enforces that, at each time $t$, the advective flux through each edge is equal to the product of the mean air velocity through that edge and the concentration at the vertex directly upwind of that edge. The advective flux term in \eqref{discretetransporteqn} is equivalent to a first-order upwind finite difference approximation of the advective flux in \eqref{conttransporteqn} within each airway. The diffusive flux term in \eqref{discretetransporteqn} is the product of the effective diffusivity \eqref{Deffdiscretedefn} in each edge and the gradient of concentration across the edge, with the gradient defined as in  \eqref{gradOperator}. Therefore, the diffusive flux in \eqref{discretetransporteqn} provides a second-order centred finite difference approximation to the diffusive flux in \eqref{conttransporteqn}. Therefore, within each airway, \eqref{discretetransporteqn} is equivalent to a finite difference approximation of the whole advection-diffusion equation \eqref{conttransporteqn}. Moreover, the formulation of \eqref{discretetransporteqn} using the discrete divergence operator \eqref{DivOperator} enforces conservation of mass of inhaled particles precisely at junctions between airways as well as at all other vertices in the network. We assume implicitly in \eqref{discretetransporteqn} that, in the acini, there is no mixing of inhaled particles from the acinar airways into the alveolar space. %\footnote{An extra note here needed referring to MPPD alveolar mixing paper. They show relatively modest effect and that effect is more significant when multiple consecutive breaths are modelled. }

It remains to specify the matrix of loss terms due to particle deposition, $\tilde{\mathsf{\Lambda}}$. We assume that loss due to particle deposition occurs via three mechanisms: inertial impaction, gravitational sedimentation and diffusion. Following, e.g., \cite{Anjilvel_1995_multiple}, we assume that deposition occurs via each mechanism independently, and we assume that the loss in a given vertex is proportional to the concentration at that vertex, so we can write
\begin{equation}
    \tilde{\mathsf{\Lambda}} = \mathrm{diag}(\,\tilde{{\lambda}}\,)\tilde{\mathsf{c}}, \quad\mbox{where}\quad \tilde{\lambda} = \tilde{\lambda}^{(\rm s)} + \tilde{\lambda}^{(\rm d)} + \tilde{\lambda}^{(\rm i)},
    \label{Lambda}
\end{equation}
with the three vectors, $\tilde{{\lambda}}^{(\rm s)}, \tilde{{\lambda}}^{(\rm d)}, \tilde{{\lambda}}^{(\rm i)}\in \mathbb{R}^{|\tilde{\mathcal{V}}|}$, representing loss due to sedimentation, diffusion and impaction, respectively. These terms are outlined in detail in §\ref{sec:deposition} below. %We use published semi-empirical or derived formulae to construct approximations for each of these terms. For impaction, we use a formula due to Zhang et al. (1997) \cite{zhang_1997_impaction} derived from computational fluid dynamics (CFD) simulations in multiple airway bifurcation geometries. For diffusion, we use a formula derived by Ingham (1991) \cite{ingham1991diffusion} which was validated against experimental results from Cohen \& Asgharian \cite{cohen1990deposition} and compares well with CFD simulations for sub-micron nanoparticles, for which impaction is not significant \cite{longest2007computational}. For sedimentation, we use a formula derived by Pich (1972) \cite{pich_1972_theory} for particles sedimenting in laminar flow through a cylindrical tube, which has been used in other models, for example by Oakes et al. \cite{oakes_aerosol_2017}. Details of how the loss terms in \eqref{Lambda} are constructed from each of these formulae are provided in \S\ref{sec:deposition} below. We do not model flow or particle deposition in the nose or oral cavity. 

Finally, we impose that the initial concentration of particles is zero everywhere in the network, $\tilde{\mathsf{c}}(t=0)=\mathsf{0}$, and we impose a boundary condition at vertex $\tilde{v}_1$,
\begin{equation} 
        D_{\mathrm{eff},1}\left(\frac{\tilde{{c}}_2 - \tilde{{c}}_1}{\tilde{l}_1}\right) =
        \left\{\begin{array}{c}
        \tilde{u}_1(\tilde{c}_1-1) \quad\mbox{for}\quad 0\leq t\leq T_b/2,\\
        0 \quad\quad\quad\quad\,\,\quad\mbox{for}\quad T_b/2< t\leq T_b.
    \end{array}\right.
    \label{mouthBCconc}
\end{equation}
where $\tilde{c}_2$ is the concentration at vertex $\tilde{v}_2$, the vertex adjacent to $\tilde{v}_1$. The boundary condition \eqref{mouthBCconc} fixes the total flux of particles into the network during inhalation, such that $\tilde{c}_1\approx1$ given the diffusive flux is small compared to the advective flux, and enforces that the diffusive flux of concentration is zero there during exhalation. To solve the equation \eqref{discretetransporteqn}, with boundary condition \eqref{mouthBCconc}, we approximate the time derivative in \eqref{discretetransporteqn} using a backwards Euler finite difference scheme, and then solve for the concentration at each time step using the BiCGSTAB solver in Eigen \cite{eigen_2010_v3}. We use a fixed time step of $\Delta t = 0.01\mathrm{s}$, after confirming that this is small enough that total deposition does not depend in any significant way on the exact value. 

\subsection{Particle deposition loss terms}\label{sec:deposition}

%\subsubsection{Gravitational sedimentation}

To model gravitational sedimentation of particles, we use a formula derived by Pich \cite{pich_1972_theory}, which is based on a simple theory of particles settling in laminar flow through a cylindrical tube, and has been used in previous deposition models \cite{oakes_aerosol_2017,poorbahrami2019regional,poorbahrami2021whole}. The formula provides a deposition efficiency, $\eta$, for a cylindrical tube, which is an approximation for the fraction of particles entering the tube that are deposited by sedimentation in the tube. Taking the same approach as in previous deposition models \cite{oakes_aerosol_2017,poorbahrami2019regional,poorbahrami2021whole}, we approximate the deposition efficiency in each edge within a given airway by $\tilde{l}_j\eta/\tilde{l}_\mathrm{a}$, where $\tilde{l}_j$ is the length of the edge and $\tilde{l}_\mathrm{a}$ is the length of the airway; the loss term associated with each edge is then $\tilde{q}_j\tilde{l}_j\eta/\tilde{l}_\mathrm{a}$, where $\tilde{q}_j$ is the flow rate through that edge. The same process is used to construct loss terms due to deposition by diffusion, except that a different formula is used for the deposition efficiency \cite{ingham1991diffusion}. We formulate the loss terms for sedimentation and diffusion below, before moving onto impaction, for which a different formulation is used since impaction is assumed to occur only at airway bifurcations. 

The loss terms due to deposition by sedimentation on all of the vertices in the network are given by
\begin{equation}
    \tilde{\lambda}^{(\rm s)} = \frac{1}{2}\tilde{\mathsf{N}}^{T}_{\rm u}{\tilde{\mathsf{V}}_{\mathrm{e}}\mathrm{diag}(\tilde{\mathsf{u}})\mathrm{diag}({\tilde{\mathsf{l}}_{\rm a}})^{-1}\boldsymbol{\eta}^{(\rm s)}},\label{sedformula}
\end{equation}
where $\tilde{\mathsf{N}}_{\rm u}$ is the unsigned incidence matrix, with components $\tilde{\mathsf{N}}_{\mathrm{u},jk} = |\tilde{\mathsf{N}}_{jk}|$, $\tilde{V}_{\mathrm{e},jj}$ is the volume of edge $\tilde{e}_j$, $\tilde{l}_{\mathrm{a},j}$ is the length of the airway containing $\tilde{e}_j$, 
\begin{equation}
    \eta^{(\rm s)}_j = \frac{2}{\pi}\left[2\epsilon_j\sqrt{1-\epsilon_j^{2/3}} - \epsilon_j^{1/3}\sqrt{1-\epsilon_j^{2/3}} + \sin^{-1}\left(\epsilon_j^{1/3}\right)\right]
    \label{etajsed}
\end{equation}
is the efficiency of deposition in the airway containing edge $\tilde{e}_j$, with 
\begin{equation}
    \epsilon_j \equiv \min\left(\frac{3u_{\rm g}\tilde{l}_{\mathrm{a},j}\sin\tilde{\theta}_jC_{\rm c}}{8\tilde{a}_j|\tilde{u}_j|},1\right),
    \label{epsilonjsed}
\end{equation}
where $\tilde{\theta}_j$ is the angle of edge $\tilde{e}_j$ with respect to the direction of gravity,
\begin{equation}
    u_{\rm g} = \frac{\rho_{\rm p}gd_{\rm p}^2}{18\mu_{\rm a}}
\end{equation}
is the settling velocity of a spherical particle in Stokes' flow, with $\rho_{\rm p}$ being the particle density and $d_{\rm p}$ the particle diameter. The operator $\tfrac{1}{2}\tilde{\mathsf{N}}_\mathrm{u}^T$ in \eqref{sedformula} assigns half of the deposition that occurs in an edge to each of the vertices connected to that edge. In \eqref{epsilonjsed}, we have enforced that $\epsilon_j\leq1$, which ensures that the deposition efficiency \eqref{etajsed} is well-defined. The definition \eqref{etajsed} ensures that the efficiency can be no greater than $1$. In all of the acinar airways, we set $\sin\tilde{\theta}_j=2/\pi$, which is equivalent to replacing $\epsilon_j$ with its expected value supposing the true gravity angles of the acinar airways are distributed uniformly in the range $[0,\pi]$.

%\subsubsection{Diffusion}

The formulation of the loss terms due to diffusion of particles is the same as in \eqref{sedformula}, except that the efficiency of deposition is based instead on a formula derived by Ingham \cite{ingham1991diffusion}. The loss terms are
\begin{equation}
    \tilde{\lambda}^{(\rm d)} = \frac{1}{2}\tilde{\mathsf{N}}^{T}_{\rm u}{\tilde{\mathsf{V}}_{\mathrm{e}}\mathrm{diag}(\tilde{\mathsf{u}})\mathrm{diag}({\tilde{\mathsf{l}}_{\rm a}})^{-1}\boldsymbol{\eta}^{(\rm d)}},\label{diffformula}
\end{equation}
where
\begin{equation}
    \eta_j^{(\rm d)} = \min\left[3.033Re_j^{-5/9}Sc^{-2/3}\left(\frac{\tilde{l}_{j}}{\tilde{a}_j}\right)^{5/9},1\right],\label{etajdiff}
\end{equation}
where $Re_j = 2|\tilde{u}_j|\tilde{a}_j/\nu_{\rm a}$ is the Reynolds number in edge $\tilde{e}_j$ and $Sc = \nu_{\rm a}/D$ is the Schmidt number. The formulation of \eqref{etajdiff} ensures that the efficiency of deposition in an airway is no greater than $1$.

%\subsubsection{Inertial Impaction}

To model inertial impaction of particles, we use a semi-empirical formula due to Zhang et al. \cite{zhang_1997_impaction}, derived using computational fluid dynamics (CFD) simulations in airway bifurcation geometries. We assume that deposition by impaction occurs at vertices representing airway bifurcations, where one parent airway branches into two daughter airways. We define a matrix $\boldsymbol{\eta}^{(\mathrm{i})}\in\mathbb{R}^{|\tilde{\mathcal{E}}|\times|\tilde{\mathcal{E}}|}$, such that $\eta_{ij}^{(\mathrm{i})}$ is the deposition efficiency due to impaction associated with flow from edge $\tilde{e}_i$ into edge $\tilde{e}_j$, given that $\tilde{e}_i$ and $\tilde{e}_j$ are adjacent and that $\tilde{e}_i$ is proximal to $\tilde{e}_j$. We define this matrix via
\begin{equation}
    \eta_{ij}^{(\mathrm{i})}  
    = \left\{\begin{array}{ll}
        \min\left[Re_i^{1/3}\left(\alpha + \beta e^{\gamma\mathrm{St}^{\delta}_i}\right)\sin\tilde{\phi}_{ij},1\right]&\quad\mbox{if $\tilde{e}_i$ is adjacent to, and proximal to, $\tilde{e}_j$} \\ 
        0 &\quad\mbox{otherwise}
        \end{array}\right.
    \label{etaijimp}
\end{equation}
where $\tilde{\phi}_{ij}$ is the branching angle between edges $\tilde{e}_i$ and $\tilde{e}_j$, 
\begin{equation}
    \mathrm{St}_i = \frac{\rho_{\rm p}d_{\rm p}^2\tilde{u}_iC_{\rm c}}{36\tilde{a}_i\mu_{\rm a}}
\end{equation}
is the Stokes number in edge $\tilde{e}_i$, $Re_i$ is the Reynolds number, and 
\begin{align}
    (\alpha,\beta,\gamma,\delta) = \begin{cases}
        (0,0.000654,55.7,0.954) \quad&\mbox{if}\quad \mathrm{St}_i<0.04,\\ 
        (0.19,-0.193,-9.5,1.565) \quad&\mbox{if}\quad \mathrm{St}_i\geq0.04,
    \end{cases}
\end{align}
are empirical parameters from Zhang et al. \cite{zhang_1997_impaction}. The loss terms at the vertices are then defined via
\begin{equation}
    \tilde{\lambda}^{(\rm i)} = |\tilde{\mathsf{N}}_-^T|\boldsymbol{\eta}^{(\mathrm{i})}\mathrm{diag}(\tilde{\mathsf{s}})\tilde{\mathsf{u}}_+,
    \label{impformula}
\end{equation}
where $\tilde{\mathsf{N}}_-$ has components $\tilde{N}_{-,jk} = \min(0,\tilde{N}_{jk})$ and $\tilde{\mathsf{u}}_+$ has components $\tilde{{u}}_{+,j} = \max(\tilde{{u}}_j,0)$. The formulation of \eqref{etaijimp}-\eqref{impformula} effectively defines a loss term associated with the flow into each daughter airway at a bifurcation, equal to the product of the deposition efficiency from \cite{zhang_1997_impaction} with the flux through that daughter airway, and sets the total loss term at the bifurcation vertex equal to the sum of the loss associated with the flow into each daughter airway. Using $\tilde{\mathsf{u}}_+$ in \eqref{impformula} effectively enforces that there is no deposition via impaction on exhalation, a common assumption, made, for example, by Oakes et al. \cite{oakes_aerosol_2017} and Anjilvel \& Asgharian \cite{Anjilvel_1995_multiple}. The formulation of \eqref{etaijimp} also enforces that the deposition efficiency is no greater than $1$.  %\textcolor{red}{In general, we find that adding impaction on exhalation makes only a minor difference to overall deposition.}\footnote{\textcolor{red}{Quantify.}} 
In the acini, we set $\sin\tilde{\phi}_{ij}=2/\pi$ at all bifurcations, which is equivalent to replacing the deposition efficiency with its average value supposing the true branching angles are distributed uniformly in the range $[0,\pi/2]$.

\begin{comment}
\subsection{Solution Methods}

We can write \eqref{discretetransporteqn} as
\begin{equation}
    \frac{\mathrm{d}\tilde{\mathsf{c}}}{\mathrm{d}t} = \mathsf{M}\tilde{\mathsf{c}},\label{discr_transp_v2}
\end{equation}
where
\begin{equation}
    \mathsf{M}(t) = -\tilde{\mathsf{V}}_{\rm v}^{-1}\tilde{\mathsf{N}}^T\mathrm{diag}(\tilde{\mathsf{s}})\left[-\mathrm{diag}(\tilde{\mathsf{u}})\tilde{\mathsf{N}}_{\rm v} + \mathsf{D}_{\rm eff}\mathrm{diag}(\,\tilde{\mathsf{l}}\,)^{-1}\tilde{\mathsf{N}}\right] - \mathrm{diag}\left(\tilde{\mathsf{\lambda}}\right)\label{Mop}
\end{equation}
To solve \eqref{discr_transp_v2}, at a given time point, $t$, we use a backwards Euler finite difference scheme to approximate the time derivative,
\begin{equation}
    \tilde{\mathsf{c}}(t+\Delta t) - \tilde{\mathsf{c}}(t) = \Delta t\mathsf{M}(t+\Delta t)\tilde{\mathsf{c}}(t+\Delta t),\label{BwdsEuler}
\end{equation}
where $\Delta t$ is a fixed time step. We use the BiCGSTAB solver in Eigen \cite{eigen_2010_v3} to solve \eqref{BwdsEuler} for $\tilde{\mathsf{c}}(t+\Delta t)$ at each time point. We use a time step of $\Delta t = 0.01\mathrm{s}$, after confirming this is small enough that total deposition does not depend on the exact value of the time step. 
\end{comment}

\section{Multiple-breath washout}\label{sec:mbw}

\begin{figure}
    \centering
    \includegraphics[width = 0.6\textwidth]{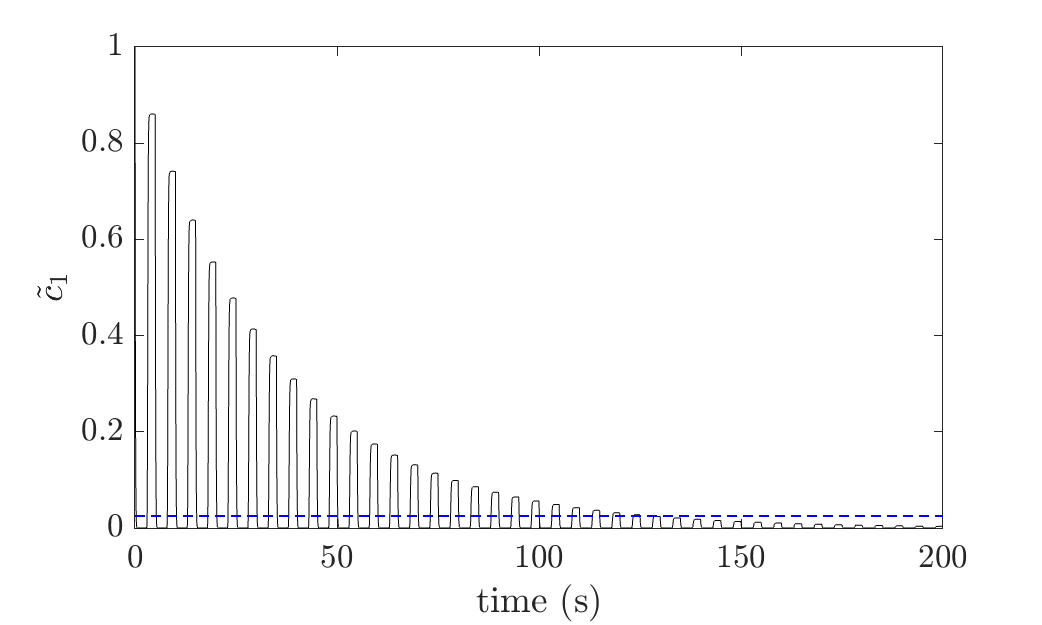}
    \caption{Concentration, $\tilde{c}_1$, of tracer gas (nitrogen) at the top of the trachea during an example multiple-breath washout simulation. The blue dashed line indicates $1/40$ of the initial concentration: when $\tilde{c}_1$ drops below this level for a whole breath, LCI is then calculated as described in §2.5 of the main text. }
    \label{fig:MBWconc}
\end{figure}

By making only minor modifications to the particle deposition model described above, we can also simulate multiple-breath washout (MBW) of an inert gas. To modify the transport equations to simulate nitrogen washout, we set the deposition loss terms to zero in equation \eqref{discretetransporteqn}, i.e. $\tilde{\Lambda}=\mathsf{0}$, and we replace the effective diffusivity \eqref{Deff} with the formula due to Scherer et al. \cite{scherer_1975_diffusivities} for gases,
\begin{equation}
    D_{\rm eff} = \left\{\begin{array}{c}
        D_{\rm tr} + 1.08\bar{u}a \quad\mbox{if}\quad \bar{u} \geq 0,\\ 
        D_{\rm tr} + 0.36\bar{u}a \quad\mbox{if}\quad \bar{u} < 0,
    \end{array}\right.
    \label{Deff2}
\end{equation}
where $D_{\rm tr}=0.225\mathrm{cm}^2\mathrm{s}^{-1}$ is the diffusivity of nitrogen in pure oxygen at $37\degree C$ \cite{verbanck2018simulation}. When solving the resulting transport equation, we solve it on another modified version of the network, $\mathcal{N}$. Unlike for the deposition model, we do not replace the acinar bags with symmetric trees of airways, instead keeping each acinus as a single vertex of the network, and assuming that the inhaled gas is always well-mixed within each acinus. This is the same approach as used by Foy et al. \cite{foy2017modelling}, and is motivated by the large diffusivity of nitrogen gas compared to the diffusivity of the aerosols we consider in the deposition model. We do discretise all of the airways in the network before solving the transport equations for MBW. We use a coarser discretisation when simulating MBW, enforcing that every airway contains at least three edges and that the maximum length of any edge is $500\mu\mathrm{m}$. This discretisation is fine enough to ensure that LCI is adequately converged: for example, for the example shown in figure \ref{fig:MBWconc}, which uses the same unconstricted lung geometry as in figure 2 in the main text, approximately doubling the number edges in the network changes the computed value of LCI by less than $0.2\%$.

We enforce that the initial concentration of nitrogen is uniform throughout the lungs before washout, $\tilde{\mathsf{c}}(t=0)=1$, and we replace the boundary condition \eqref{mouthBCconc} with
\begin{align} 
        D_{\mathrm{eff},1}\left(\frac{\tilde{{c}}_2 - \tilde{{c}}_1}{l_1}\right) =
        \begin{cases}
        \tilde{u}_1\tilde{c}_1 \,\quad&\mbox{for}\quad 0\leq t\leq nT_b/2,\\
        \quad 0\quad\quad&\mbox{for}\quad nT_b/2< t\leq nT_b,
    \end{cases}
    \label{MBWmouthBC}
\end{align}
for $n = \{1,2,\dots,n_b\}$, where $n_b$ is the total number of breaths simulated, which enforces that the flux of concentration into the network is zero on inhalation. The number of breaths, $n_b$, is chosen to be large enough that LCI can be calculated as described in §2.5 of the main text. We solve the resulting system of equations in the same way as described above. An example output, showing the concentration of tracer gas at the top of the trachea, is illustrated in figure \ref{fig:MBWconc}.

\section{Gamma scintigraphy}

\begin{figure}
    \centering
    \includegraphics[width = \textwidth]{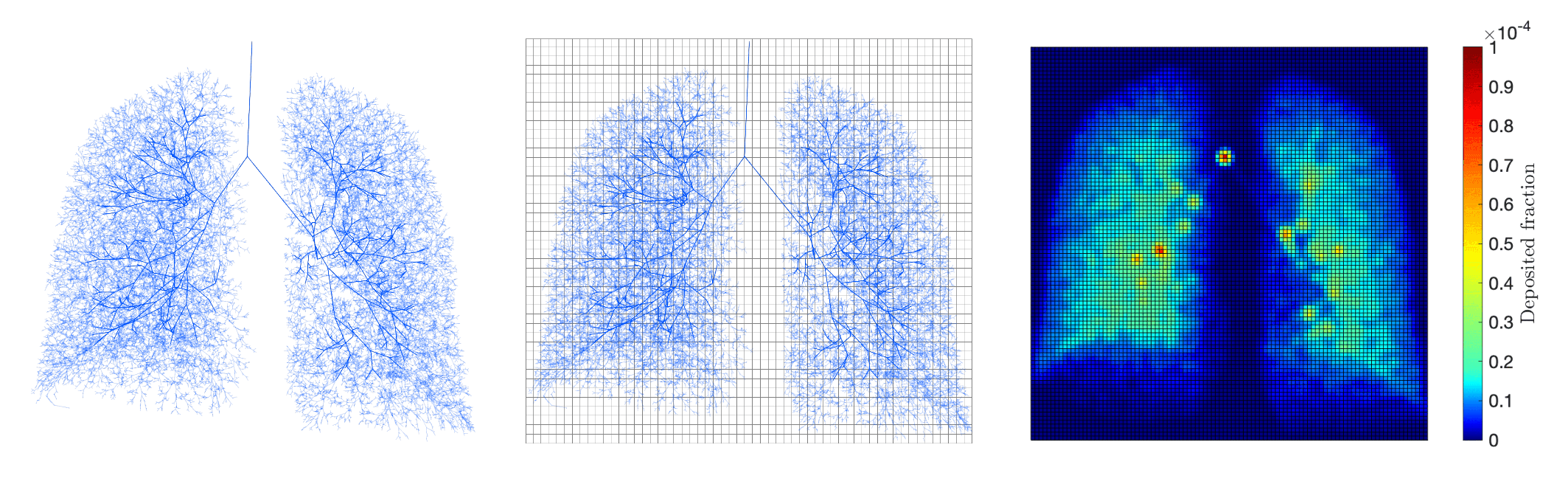}
    \caption{Schematic showing the process of generating a scintigraphy image. The 3D lung network is oriented front-on (left). The 2D plane is then divided into many equally sized pixels (centre). For each pixel, deposition occurring at every vertex of the network lying within that pixel is summed. This data is then converted into a scintigrpahy image (right), using kernel density estimation.}
    \label{fig:scint}
\end{figure}

The following process is used to generate a simulated gamma scintigraphy image, given the output from a particle deposition simulation. The 3D lung network is oriented front-on in the 2D plane (see figure \ref{fig:scint}). We then divide the 2D plane surrounding the network into a grid of many equally sized pixels, as illustrated schematically in figure \ref{fig:scint}. We will call the pixel $i^{\rm th}$ from the left edge of the grid and $j^{\rm th}$ from the bottom $p_{ij}$, and we will say the position of the centre of that pixel is $(x_{ij},y_{ij})$. %To do so, we define a grid spacing, $\Delta x$, in the $x$-direction, and a grid spacing, $\Delta y$, in the $y$-direction, in the plane. 
We define the grid so that the left-most vertex of the network lies within the left-most column of pixels, the right-most vertex lies within the right-most column of pixels, the lower-most vertex of the network lies within the bottom row of pixels and the upper-most vertex of the network lies within the top row of pixels. We take the pixels to be square, with height and width $0.74\mathrm{mm}$, which for the plots shown in the main text (and figure \ref{fig:scint}) results in approximately $150,000$ pixels. Then, for each pixel, we sum the deposited fraction from all of the network's vertices that lie within that pixel. The spatial orientation of the acinar airways is not explicitly defined in the lung networks, so deposition that occurs in acinar airways is assigned to the vertex at the distal end of the terminal bronchiole to which that acinus is attached. Summing the depositions gives, for each pixel $p_{ij}$, a corresponding value, $M_{ij}$, of the total deposited fraction in $p_{ij}$. We then use kernel density estimation to smooth this deposited fraction data. We assign an integer value to each pixel, which approximates its deposited fraction: we define $\hat{M}$, such that $\hat{M}_{ij}=\lfloor10^6\times M_{ij}\rfloor$, i.e., we add one to $\hat{M}_{ij}$ for each multiple of one millionth of the total inhaled dose deposited in $p_{ij}$. This can be interpreted as an assumption that, if there are one million discrete particles inhaled in total, then $\hat{M}_{ij}$ is approximately the number of particles that are deposited in pixel $p_{ij}$. Then, we form a list of coordinate positions, $\boldsymbol{X}$, such that each entry is a coordinate position of the centre of some pixel. For each pixel $p_{ij}$, $\boldsymbol{X}$ is assigned exactly $\hat{M}_{ij}$ entries with value $(x_{ij},y_{ij})$. Therefore, if we again suppose there were one million inhaled particles, $\boldsymbol{X}$ can be interpreted as a list of the spatial position where each particle deposited. Finally, we apply \textsc{Matlab}'s kernel density estimation function, \texttt{ksdensity}, to the list of points, $\boldsymbol{X}$, with a bandwidth of $4.0$, which defines the degree to which the data is smoothed. An example of the simulated scintigraphy image is shown in figure \ref{fig:scint}.

\printbibliography